\documentclass[apj]{emulateapj}
\usepackage{amsmath}
\usepackage{apjfonts}
\usepackage{multirow}
\usepackage{hyperref}
\usepackage{array}
\newcommand{\DDir}{\relax{D\kern-.7em{/}}}

\newcommand{\xra}{\xrightarrow}

\newcommand{\be}{\begin{equation}}
\newcommand{\ee}{\end{equation}}
\newcommand{\bea}{\begin{equation*}}
\newcommand{\eea}{\end{equation*}}

\newcommand{\pr}{\partial}

\newcommand{\nin}{\relax{\in\kern-.8em{/}}}

\newcommand{\bt}{\beta}

\newcommand{\lm}{\lambda}

\newcommand{\sig}{\sigma}

\newcommand{\vt}{\textrm{v}}

\newcommand{\cm}{\mbox{cm}}

\newcommand{\se}{\mbox{s}}
\newcommand{\yr}{\mbox{yr}}

\newcommand{\erg}{\mbox{erg}}

\newcommand{\Mpc}{\mbox{Mpc}}
\newcommand{\eV}{\mbox{eV}}
\newcommand{\keV}{\mbox{keV}}

\newcommand{\gr}{\mbox{g}}

\newcommand{\sref}{\S~\ref}

\begin{document}
\title{Non-relativistic radiation mediated shock breakouts: \\III. Spectral properties of SN shock breakout}
\author{Nir Sapir\altaffilmark{1}, Boaz Katz\altaffilmark{2,3} and Eli Waxman\altaffilmark{1}}

\altaffiltext{1}{Dept. of Particle Phys. \& Astrophys., Weizmann Institute of Science, Rehovot 76100, Israel}
\altaffiltext{2}{Inst. for Advanced Study, Princeton, NJ 08540, USA}
\altaffiltext{3}{John Bahcall Fellow, Einstein Fellow}

\begin{abstract}
The spectrum of radiation emitted following shock breakout from a star's surface with a power-law density profile $\rho \propto x^n$ is investigated. Assuming planar geometry, local Compton equilibrium and bremsstrahlung emission as the dominant photon production mechanism, numerical solutions are obtained for the photon number density and temperature profiles as a function of time, for hydrogen-helium envelopes. The temperature solutions are determined by the breakout shock velocity $v_0$ and the pre-shock breakout density $\rho_0$, and depend weakly on the value of $n$. Fitting formulas for the peak surface temperature at breakout as a function of $v_0$ and $\rho_0$ are provided, with $\mathcal{T}_{\rm peak}\approx 9.44\exp{[12.63(v_0/c)^{1/2}]}~\eV$, and the time dependence of the surface temperature is tabulated. The time integrated emitted spectrum is a robust prediction of the model, determined by $\mathcal{T}_{\rm peak}$ and $v_0$ alone and insensitive to details of light travel time or slight deviations from spherical symmetry. Adopting commonly assumed progenitor parameters, breakout luminosities of $\approx10^{45}~\erg~\se^{-1}$ and $\approx10^{44}~\erg ~\se^{-1}$ in the 0.3-10 keV band are expected for BSG and RSG/He-WR progenitors respectively ($\mathcal{T}_{\rm peak}$ is well below the band for RSGs, unless their radius is $\sim 10^{13}~\cm$). $>30$ detections of SN1987A-like (BSG) breakouts are expected over the lifetime of \textit{ROSAT} and \textit{XMM-Newton}. An absence of such detections would imply that either the typical parameters assumed for BSG progenitors are grossly incorrect or that their envelopes are not hydrostatic.
The observed spectrum and duration of XRF 080109/SN2008D are in tension with a non-relativistic breakout from a stellar surface interpretation.
\end{abstract}
\keywords{radiation hydrodynamics --- shock waves --- supernovae: general}

\section{Introduction}\label{sec:Introduction}
In a supernova (SN) explosion, a radiation mediated shock (RMS) traverses the stellar envelope. When the shock breaks out from the surface of the star, radiation is emitted with a prompt bright flash. Observations of breakouts provide important information about the progenitor. A detailed understanding of the expected bolometric and spectral properties is required for an accurate estimate of the progenitor's radius and for planning of transient search strategies.

The bolometric and hydrodynamical properties of a non-relativistic planar RMS breaking out from a stellar envelope have been studied in \citet{Sapir11} (hereafter Paper I). The analysis was preformed by neglecting matter thermal energy and approximating radiation energy transport as diffusion with constant opacity $\kappa$. The bolometric light curves and the hydrodynamic profiles were solved and fitted by analytical formulas. The observed properties of SN shock breakouts were derived in \citet{Katz12} (hereafter Paper II) using the planar results, assuming spherical symmetry and taking into account limb darkening.

The spectral energy distribution of the emission is dependent on the interaction of radiation and matter.
In local thermal equilibrium (LTE), the radiation temperature is directly determined from the radiation energy density, and thus from the hydrodynamic solution.
However, in general the radiation spectrum is determined by processes of absorption, emission, scattering and adiabatic compression.
As these processes have different time scales, the radiation state is time-dependent, and it may not be in LTE during shock propagation in the envelope.
For instance, in fast RMSs there is not enough time for photons to be produced or absorbed in the shock front, and the radiation will reach LTE only in the far downstream \citep{Weaver76,Katz10}.

Earlier work studied the spectral properties of radiation emitted following breakout, either with numerical calculations of specific progenitor structures \citep{Klein78,Ensman92,Blinnikov98,Blinnikov00,Tolstov10} or by analytical estimates \citep{Katz10,Nakar10}.
Past numerical work, however, either assumed LTE or used single-group approximations or multi-group calculations without fully accounting for the effect of Compton scattering on the spectral energy distribution.
In particular, these works ignored incoherent Compton Scattering, in which a photon changes its energy upon scattering on an electron.
This effect plays a key role in the problem, and drives radiation and matter into a Compton equilibrium.
Consequently, the radiation temperature can be orders of magnitude higher compared to the LTE case \citep{Weaver76,Katz10}.
Past analytical work applied either steady state solutions to the time dependent problem or assumed only local production of photons, not accounting for the role of diffusion in determining the emitted spectrum.

This work follows the effective photon approximation used in the temperature calculation of steady state RMSs \citep{Weaver76, Katz10}, and applies it to solve the time dependent problem of planar shock breakout from a stellar envelope.
The problem is solved under the assumptions of local Compton equilibrium and bremsstrahlung as the dominant emission rate.
Solutions are provided for the temperature and the emitted radiation spectrum.

This paper is organized as follows. The formulation of the problem and the governing equations are presented in \sref{sec:Formulation of the problem}. Results for the time dependent profiles of the temperature and for the surface temperature are described in \sref{sec:Results}. The surface temperature is then used to estimate the emitted spectrum in \sref{sec:EmisSpec}. The expected emission from possible progenitors and the implied detection rates are discussed in \sref{sec:Shock breakout searches}. Comparisons with previous results in the literature and with the observations of the X-ray transient 080109/SN2008D are given in \sref{sec:Previous works and observations}. The main conclusions are summarized in \sref{sec:Conclusions}.

\section{Formulation of the problem}\label{sec:Formulation of the problem}
Consider the problem of a planar non relativistic RMS propagating in a stellar envelope, and breaking out from its surface. Assuming the energy density is dominated by radiation, and the opacity is dominated by Thomson scattering, the hydrodynamic problem can be described with flow conservation equations coupled to a diffusion equation with constant opacity, and was solved in Paper I.
In order to solve for the radiation temperature, further assumptions are required: here we solve for the temperature assuming the radiation is in local Compton equilibrium and the emission mechanism is bremsstrahlung emission,

The initial density profile is assumed to be a power-law of the distance from the surface, $\rho\propto x^n$.
In terms of the optical depth with respect to the surface, $\tau=\kappa\int \rho dx$, the density profile is given by
\begin{equation}\label{eq:InitialProfileRho}
\rho(\tau)=\rho_0 (\bt_0\tau)^{n/(n+1)},
\end{equation}
where $\bt_0=v_0/c$.
At large optical depth, the asymptotic shock velocity can be described by the pure hydrodynamic solution \citep{Sakurai60}, and parameterized by
\begin{equation}\label{eq:InitialProfileV}
v_{\rm sh}\xra[\bt_0\tau>>1]{}v_0 (\bt_0\tau)^{-\lm/(n+1)}.
\end{equation}
The values of $\rho_0$ and $v_0$ are the density and velocity at the point $\tau=c/v_{\rm sh}$, which is obtained in a pure hydrodynamic shock propagation.

In Compton equilibrium, the radiation is characterized by a Wien spectrum, $de/d\nu \propto \nu^3 {\rm e}^{-\nu/T_\gamma}$, with radiation temperature, $T_\gamma$, given by the electron temperature, $T_\gamma=T$.
The Wien spectrum describes the solution for saturated Comptonization of soft photons, consistent with the effective photon approximation which is applied in this work \citep{Blandford81}.
Accordingly, the relation between radiation temperature, pressure and photon number density is
\begin{equation}\label{eq:ComptonTemp}
T_\gamma=\frac{p}{n_{\gamma}}.
\end{equation}
Note that in the case of LTE $T\approx0.9 p/n_{\gamma}$, and eq.~\eqref{eq:ComptonTemp} introduces a $10\%$ error in the radiation temperature.

The number density of photons is determined by photon transport in space and photon production by the plasma, which depends on the plasma temperature and density.
As the hydrodynamic flow equations are independent of temperature, the detailed interaction of photons with matter does not affect the hydrodynamics.
The independent solution for the energy density and density profiles, found in Paper I, can therefore be used to calculate the photon number density and temperature profiles as a function of time.

\subsection{Hydrodynamics and energy diffusion equations}\label{sec:Hydrodynamics}
We briefly describe the hydrodynamic equations which are given in Paper I.
Lagrangian coordinates are used to describe the flow. The mass coordinate $m$ equals the mass per unit area enclosed between each mass element and the surface. The equations describing the position, $x$, velocity $v$ and energy density $e$ as a function of $m$ and time $t$ are
\begin{align}
&\pr_t x=v, \label{eq:hydro1} \\
&\pr_{t} v=-\pr_mp,  \label{eq:hydro2}\\
&\pr_t(e/\rho)=-\pr_m j_e-p\pr_m v, \label{eq:hydro3}
\end{align}
where $\rho=\pr_xm$ is the density, $j_e=-c(3\kappa)^{-1}\pr_me$ is the energy flux (energy current density), $p=e/3$ is the radiation pressure and $\kappa$ is the constant opacity. The boundary conditions at the surface are taken as
\begin{equation}\label{eq:SurfaceBc}
e(\bt_0\tau=0)=0,~~~\pr_t v(\bt_0\tau=0)=\frac{\kappa}{c}j_e(\bt_0\tau=0).
\end{equation}
The dimensional parameters $\kappa, \rho_0$ and $v_0$ along with the dimensionless parameter $n$ completely define the hydrodynamic problem. The hydrodynamic profiles can be obtained, solving the equations of motion \eqref{eq:hydro1}-\eqref{eq:hydro3}, using the initial density profile \eqref{eq:InitialProfileRho}, the asymptotic shock velocity \eqref{eq:InitialProfileV} and the surface boundary conditions \eqref{eq:SurfaceBc} (Paper I).

\subsection{Photon number density and temperature}\label{sec:Photon number density}

In the effective photon approximation the photon number density is calculated from the photon emission rate, corrected for absorption \citep{Weaver76}.
Although the photon emission rate diverges logarithmically at low energies, photons that do not up-scatter in energy are absorbed, and a lower cutoff energy can be introduced.
This cutoff energy is the lowest energy from which a photon can be "thermalized" to the plasma temperature $T$.

The transport of photons is modeled by a diffusion equation with an additional source term, $Q_{\gamma}$, which accounts for the local net production rate of photons,
\begin{equation}\label{eq:PhotonNumberEq}
\pr_{t}(n_{\gamma}/\rho)=-\pr_m j_{\gamma}+Q_{\gamma},
\end{equation}
where the photon number current is
\begin{equation}\label{eq:PhotonCurrent}
j_{\gamma}=-\frac{c}{3\kappa}\pr_m n_{\gamma}.
\end{equation}
Note that Compton scattering conserves the number of photons, and therefore scattering does not directly appear in the equation for the photon number density.

At large optical depth the radiation is found at LTE.
The inner boundary condition for the photon number current is accordingly taken as
\begin{align}\label{eq:InnerBcnga}
j_{\gamma}(\bt_0\tau\rightarrow\infty)=&\frac{1}{4}\left[\frac{e(\bt_0\tau\rightarrow\infty)}{0.9a_{\rm BB}}\right]^{-1/4}j_e(\bt_0\tau\rightarrow\infty),
\end{align}
where $a_{\rm BB}=(\hbar c)^{-3}\pi^2/15$ is the Stefan-Boltzmann energy density coefficient and the $0.9$ factor accounts for the error introduced using eq.~\eqref{eq:ComptonTemp} in LTE.
At the surface, the photon number density boundary condition is taken similar to the energy density boundary condition,
\begin{equation}\label{eq:SurfaceBcnga}
n_{\gamma}(\bt_0\tau=0)=0.
\end{equation}
Note that the boundary conditions at the surface are taken as zero energy and zero photon number density (see also Paper I), but the photon temperature is finite.

The source term in eq.~\eqref{eq:PhotonNumberEq} is the net photon production rate per unit mass, which depends both on temperature and density of the plasma,
\begin{equation}\label{eq:PhotonGeneratioRate}
Q_{\gamma}(\rho,T)=\frac{1}{\rho}q_{\gamma}^{\rm eff}(\rho,T)f_{\rm abs}(\rho,T),
\end{equation}
where $q_{\gamma}^{\rm eff}$ is the photon number density emitted per unit time and $f_{\rm abs}$ is a dimensionless factor which accounts for absorption.

In the temperature range $10 ~\eV < T < 50 ~\keV$, the dominant mechanism for photon production in a fully ionized hydrogen-helium plasma is electron-ion bremsstrahlung emission, while pair production and free-bound emission are negligible.
At the high-temperature end, double Compton emission, which is not included in our calculations, is subdominant but not entirely negligible compared to bremsstrahlung emission (see appendix A). Its inclusion will introduce $\sim10\%$ corrections to the derived peak temperature at high shock velocities, $v_{\rm sh}/c>0.3$
Consider a completely ionized plasma with atomic number $Z=\Sigma_i x_i Z_i$ and atomic weight $A=\Sigma_i x_i A_i$, where $x_i$ are the atomic fractions of each element in the envelope composition.
The effective photon emission rate is given by \citep{Weaver76,Katz10}
\begin{equation}\label{eq:PhotonGeneratioRateCB}
\begin{split}
q_{\gamma}^{\rm eff}(\rho,T)=& \frac{2^{3/2}}{\pi^{3/2}}\frac{\Sigma_i x_i Z_i^2 \Sigma_i x_i Z_i}{\left(\Sigma_i x_i A_i\right)^2} \\
& \times \alpha_{\rm e} \sig_{\rm T} c \left(\frac{\rho}{m_{\rm p}}\right)^2 \left(\frac{m_{\rm e} c^2}{T}\right)^{1/2} \Lambda_{\rm g,eff}(\rho,T).
\end{split}
\end{equation}
where $\sigma_{\rm T}$ is the Thomson cross section and the factor $\Lambda_{\rm g,eff}$ expresses the integral of the bremsstrahlung emission spectrum down from a low energy cutoff $\nu_{\rm c}$ (as this integral diverges logarithmically).
The integral over the emission spectrum, including the gaunt factor calculated in the Born approximation, can be described as a product of two separate functions, $\Lambda_{\rm g,eff}=g_1(\lambda)E_1(\lambda)$, where $\lambda =\nu_{\rm c}/T$ \citep{Weaver76}.
These functions can be calculated numerically.
The function $g_1(\lambda)$ is approximated by
\begin{equation}
g_1(\lambda)=1.226-0.475\log(\lambda)+0.0013\log^2(\lambda),
\end{equation}
and $E_1(\lambda)$ is the exponential integral function
\begin{equation}
E_1(\lambda)=\int_\lambda^\infty \frac{{\rm e}^{-x}}{x}dx,
\end{equation}
which can be approximated as $E_1(\lambda)=-\log(\lambda)-0.5772$ for small $\lambda$.

The cutoff energy $\nu_{\rm c}$ is determined by the lowest energy of emitted photons that reach thermalization with the plasma by up-scattering, before being absorbed.
The energy of photons which up-scatter $T/(m_{\rm e} c^2)$ times before getting absorbed is given by \citep{Chapline73,Weaver76},
\begin{equation}\label{eq:nuc1}
\begin{split}
\nu_{{\rm c}}=&\left(\frac{8\pi^3}{3}\right)^{1/4}\alpha_{\rm e}^{-1} \left(\frac{\Sigma_i x_i Z_i^2}{\Sigma_i x_i A_i}\right)^{1/2} (\rho m_{\rm p}^{-1} r_{\rm e}^3)^{1/2} \left(\frac{T}{m_{\rm e} c^2}\right)^{-5/4} m_{\rm e} c^2 \\
=&1.88 \left(\frac{\Sigma_i x_i Z_i^2}{\Sigma_i x_i A_i}\right)^{1/2} \rho_{-9}^{1/2} T_{\keV}^{-5/4}~\eV.
\end{split}
\end{equation}
where $\rho=10^{-9} \rho_{-9}~\gr~\cm^{-3}$ and $T=1 T_{\keV}~\keV$.
When the Compton y-parameter is lower than unity, $4T/(m_{\rm e}c^2)\tau^2<1$, photons escape the medium before being up-scattered in energy.
Therefore, $\Lambda_{\rm g,eff}$ has a minimal value of unity.

In order to account for absorption, the following term is introduced in the effective emission rate \citep{Weaver76},
\begin{equation}\label{eq:AbsCorr}
f_{\rm abs}=1-\left(\frac{T_{\rm eq}}{T}\right)^4,
\end{equation}
where the equilibrium temperature is calculated from the local energy density,
\begin{equation}\label{eq:Tblackbody}
T_{\rm eq}=\left(\frac{e}{0.9 a_{\rm BB}}\right)^{1/4}.
\end{equation}
This ensures that there is no net photon production when the radiation is in LTE with matter.

To conclude, the photon temperature and number density as a function of time are described by the photon diffusion equation \eqref{eq:PhotonNumberEq} with the boundary conditions \eqref{eq:InnerBcnga} and \eqref{eq:SurfaceBcnga}, and using the Compton equilibrium relation \eqref{eq:ComptonTemp}, the effective photon generation rate \eqref{eq:PhotonGeneratioRate} and the hydrodynamic profiles that are obtained independently (see \sref{sec:Hydrodynamics}).

\subsection{Dimensionless parameters}\label{sec:Dimensionless parameters}
In the considered temperature range, $10 ~\eV < T < 50 ~\keV$, and density range, $\rho < 10^{-6} ~\gr~\cm^{-3}$, opacity is dominated by electron scattering \citep{Cox68}.
The total opacity is then $\kappa=(\Sigma_i x_i Z_i/\Sigma_i x_i A_i) (\sigma_{\rm T}/m_{\rm p})$.
Comparing the radiation energy density emitted in a dynamical time to the shock energy density, $q_{\gamma}^{\rm eff} T t_d=\rho v_0^2$, the following photon temperature scale is obtained
\begin{equation}\label{eq:TempScale}
T_0=\left(\frac{\Sigma_i x_i A_i}{\Sigma_i x_i Z_i^2}\right)^2\bt_0^8 m_{\rm p} c^2.
\end{equation}
Introducing dimensionless variables, $\tilde a=a/a_0$, with
\begin{equation}\label{eq:A0}
t_0=\frac{c}{\kappa\rho_0 v_0^2},~~~ e_0=\rho_0 v_0^2,~~~n_{\gamma,0}=e_0/T_0, ~~~ m_0=(\kappa\bt_0)^{-1},
\end{equation}
eq.~\eqref{eq:PhotonNumberEq} can be re-written as
\begin{equation}\label{eq:PhotonNumberEqDimLess}
\pr_{\tilde t}\tilde n_{\gamma}=\frac{\tilde n_\gamma}{\tilde \rho}\pr_{\tilde t} \tilde \rho+\frac{\tilde \rho}{3}\pr_{\tilde m}^2 \tilde n_{\gamma}+\tilde{q}_{\gamma}^{\rm eff}\left(1-\frac{\tilde e}{\tilde T^4}\frac{T_{{\rm eq},0}^4}{T_0^4} \right),
\end{equation}
with a dimensionless emission rate
\begin{equation}
\tilde{q}_{\gamma}^{\rm eff}(\rho,T)=\frac{2^{3/2}}{\pi^{3/2}}\alpha_{\rm e}\left(\frac{m_{\rm e}}{m_{\rm p}}\right)^{1/2}\left(\frac{\rho}{\rho_0}\right)^2 \left(\frac{T}{T_0}\right)^{-1/2}\Lambda_{\rm g,eff}(\lambda)
\end{equation}
and equilibrium temperature
\begin{equation}\label{eq:EquilibriumTemp}
T_{{\rm eq},0}=\left(\frac{\rho_0 v_0^2}{0.9 a_{\rm BB}}\right)^{1/4}.
\end{equation}
Note that $\nu_{{\rm c}}$, and therefore also $\Lambda_{\rm g,eff}$, scales with $\Sigma_i x_i Z_i^2/\Sigma_i x_i A_i$.

For non-relativistic velocities, a single dimensionless parameter appears in the hydrodynamic problem, namely, the power-law index $n$.
Thus, for a given index $n$, the hydrodynamics follow a unique solution which is scalable with the parameters $\kappa,\rho_0$ and $v_0$ (see Paper I).
However, the temperature solution does not generally produce such a universal behavior, as there are two more dimensionless parameters in the problem.

Eq.~\eqref{eq:PhotonNumberEqDimLess} implies the existence of three interesting regimes. At LTE the temperature behavior is directly determined by the universal hydrodynamic solution, and therefore $T/T_{{\rm eq},0}$ exhibits a universal behavior. When $\nu_{\rm c} \gtrsim 0.45 T$, $\Lambda_{\rm g,eff}$ is at its lower limit of unity, and there are two energy scales in the problem in addition to the shock energy. The first is the equilibrium temperature $T_{{\rm eq},0}$ and the second is the temperature obtained in bremsstrahlung emission over a dynamical time, $T_0$.
For a given index $n$, the photon number density and temperature in this regime depend on the dimensionless parameter $T_0/T_{\rm eq,0}\propto \rho_0^{-1/4}\bt_0^{15/2}$.
When $T \gg T_{{\rm eq},0}$ there are again two energy scales in the problem (in addition to the shock energy), $T_0$ and the lower cutoff energy $\nu_{\rm c}$, and the solution is determined by 2 dimensionless parameters, $n$ and $T/\nu_{\rm c}\propto \rho_0^{-1/2}\bt_0^{18}$.

\section{Results}\label{sec:Results}
The photon diffusion equation, eq.~\eqref{eq:PhotonNumberEq}, was numerically solved using an implicit scheme in time, on a uniform mass grid. The hydrodynamic equations, eqs. \eqref{eq:hydro1}-\eqref{eq:hydro3} were solved simultaneously, in a manner described in Paper I. In order to achieve fast convergence, the initial photon number density was derived from the initial energy density assuming LTE, $n_{\gamma}=(0.9a_{\rm BB})^{1/4}e^{3/4}/3$. The inner boundary condition, eq.~\eqref{eq:InnerBcnga}, was obtained from Sakurai's time dependent hydrodynamical solution \citep{Sakurai60}.

Note that in the diffusion model the radiation field instantaneously spreads to the far upstream. The plasma and the radiation field at the far upstream quickly reaches LTE, so that a finite number of photons is advected from the upstream to shock front. However, since the resulting energy density at the far upstream is small, the number of photons advected to the shock front is insignificant compared to the number produced at the shock's immediate downstream.

The resulting profiles of the temperature as a function of optical depth at different times are given in \sref{sec:TempProfiles} for a hydrogen-helium envelope with $n=3$. The time dependence of the surface temperature is described in \sref{sec:SurfTemp}. Note that as calculated temperatures scale with atomic number and weight as $\Sigma_i x_i Z_i^2/\Sigma_i x_i A_i$ (see eq.~\eqref{eq:TempScale}), the following results are applicable to envelopes with any hydrogen-helium composition.

\subsection{Temperature profiles}\label{sec:TempProfiles}

\begin{figure}[h]
\includegraphics[scale=0.8]{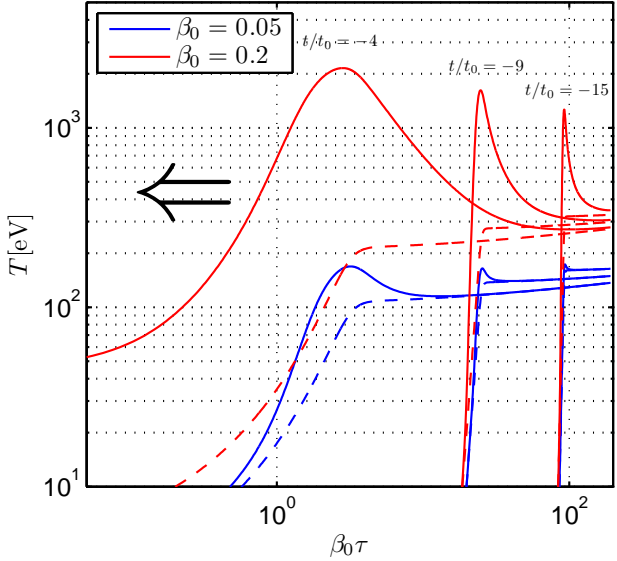}
\caption{Temperature profiles as a function of optical depth at different times prior to breakout (solid lines). Also plotted are the LTE temperature profiles (dashed line). $\bt_0c$ is the shock velocity "at breakout", i.e. the velocity for which
$\tau=\bt_0^{-1}$ in the Sakurai solution, and $t_0$ is defined in eq.~\eqref{eq:A0}. \label{fig:T_vs_M}}
\end{figure}

\begin{figure}[h]
\includegraphics[scale=0.8]{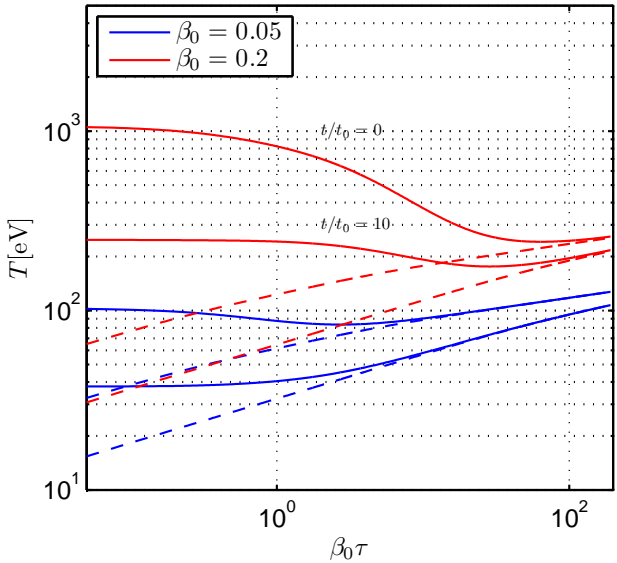}
\caption{Temperature profiles as a function of optical depth at different times following breakout (solid lines). Also plotted are the black-body temperature profiles from eq.~\eqref{eq:Tblackbody} (dashed line). \label{fig:T_vs_M_2}}
\end{figure}

Temperature profiles at different times during shock propagation, calculated with an ion number density, $n_{\rm p,0}=\rho_0/(Am_{\rm p})$, of $n_{\rm p,0}=10^{15}A^{-1}~\cm^{-3}$ and different values of $\bt_0$ are shown in figure \ref{fig:T_vs_M}.
Temperature profiles at different times following breakout, calculated with different values of $\bt_0$ and $\rho_0$ are presented in figure \ref{fig:T_vs_M_2}.
Also presented (in both figures) are the black-body temperature profiles, calculated from the energy density profiles using eq.~\eqref{eq:Tblackbody}.
As can be seen, temperatures in excess of $2~\keV$ are obtained for $\bt_0=0.2$.
In addition, even for a low shock velocity, $\bt_0=0.05$, the temperature at an optical depth of $\bt_0^{-1}$ is a factor 2 higher than the LTE value.

There are several competing effects determining the temperature at the shock front, which is affected both by photons produced locally at the shock front and by photons diffusing from the downstream. Diffusion of photons from the far downstream lowers the radiation temperature, as there are more photons to share the shock's energy. The number of photons required to achieve thermal equilibrium decreases as the shock propagates outwards and its energy density decreases. On the other hand, bremsstrahlung photon emission decreases as the shock propagates outwards to lower densities, and the optical depth of the post-shock shell from which photons may diffuse up to the shock, $\Delta \tau =7\bt_{\rm sh}^{-1}$, is decreasing as the shock accelerates.

Note that since photons produced up to an optical depth of $\Delta \tau =7\bt_{\rm sh}^{-1}$ in the downstream diffuse and reach the shock front while being up-scattered in energy, the y-parameter relevant for Comptonization at breakout is roughly $\approx 200 \bt_{\rm 0}^{-2}(T/m_{\rm e}c^2)\gg 1$, far exceeding unity also at optical depths $\tau<(m_{\rm e}c^2/4T)^{1/2}$.

LTE is obtained at the shock front when there is enough time for photons to be produced at the front, or reach it by diffusion from the downstream. In the far downstream the radiation is at LTE, but as the shock traverses the gas, the radiation is driven out of thermal equilibrium at the shock front. This occurs also for low shock velocities, in which LTE is naively expected to hold even close to breakout. In a steady state shock, mass elements in the downstream move with the same velocity, and the shock front is affected by photons diffusing from an optical depth of $7 \bt_0^{-1}$ into the downstream. However, as the shock gets closer to the star's surface the matter velocity gradient steepens, and matter at the shock front moves faster than matter in the downstream. As a result, the shock front "sees" a smaller optical depth into the downstream than in the steady-state case.

It is possible to roughly estimate the range of $\bt_0$ and $\rho_0$ for which LTE is achieved from the steady-state RMS. \citet{Weaver76} concludes that the shock front of a steady-state RMS would be in LTE for shock velocities $\bt_0<0.02 (\rho_0/10^{-9}~\gr~\cm^{-3})^{1/30}$, based on the effective photon approximation also employed in this work.
\citet{Nakar10} conclude that the shock front would be in LTE for shock velocities $\bt_0<0.05 (\rho_0/10^{-9}~\gr~ \cm^{-3})^{1/30}$, based on local production of photons in the immediate downstream.
Our results show that the estimate provided by \citet{Weaver76} is more accurate for the shock breakout problem.

Analytical estimates for the maximal temperature at the shock front can be obtained by applying the steady-state RMS temperature solution to Sakurai's hydrodynamical solution.
Taking the initial density and shock velocity values as a function of optical depth from eqs. \eqref{eq:InitialProfileRho}-\eqref{eq:InitialProfileV}, and using them as a uniform input for an RMS propagating in a homogeneous medium \citep{Weaver76,Katz10}, allows one to estimate the maximal temperature as a function of optical depth.
A comparison with the steady-state estimate reveals good agreement at large optical depth, but an increasing discrepancy as the shock approaches the surface.
At an optical depth of $\tau=\beta_0^{-1}$ the calculated temperature may a factor 2-5 lower than the analytical estimate.
This difference originates from the large photon number density in the downstream, compared to the steady-state case.
Thus, more photons arrive from the downstream, lowering the temperature at the shock front.
As the shock accelerates and radiation is driven away from equilibrium, the effect becomes more pronounced.
Hence, calculating the emitted spectrum from the steady-state estimate may produce large errors.

\subsection{Surface temperature}\label{sec:SurfTemp}
The observed emission is determined by photons escaping from the surface of last scattering, which is at an optical depth of $\tau\approx 1$.
The emitted radiation spectral energy distribution can be described using the surface temperature, $\mathcal{T}(t)=T(\bt_0\tau \rightarrow 0,t)$, in the limit $\beta_0 \ll 1$.
When Thomson opacity dominates over bremsstrahlung opacity, radiation is not at LTE at this optical depth and the surface temperature does not equal the effective temperature, $T_{\rm eff}=[4\mathcal{L}/(a_{\rm BB}c)]^{1/4}$, where $\mathcal{L}$ is the bolometric emitted flux.

In the case of scattering-dominated opacity the temperature at the surface is very close to the temperature at an optical depth of $\tau=\beta_0^{-1}$.
Photons arriving from an optical depth $\tau=\bt_0^{-1}$ scatter on their way to the surface, but almost no new photons are produced.
As $n_\gamma\gg Z n_{\rm p}$, the electrons' temperature is expected to be equilibrated with the arriving photons, and if Compton scattering dominates over emission and absorption the plasma and the radiation reach Compton equilibrium.
Therefore, local Compton equilibrium exists even at optical depths at which the Compton y-parameter is lower than unity.
Additionally, photons with energies $<\bt_0^{2} m_{\rm e}c^2$ do not lose significant energy diffusing from $\tau=\beta_0^{-1}$ to $\tau=1$, and the energy required to raise the electrons' temperature to the radiation temperature is small.

\begin{figure}[h]
\includegraphics[scale=0.8]{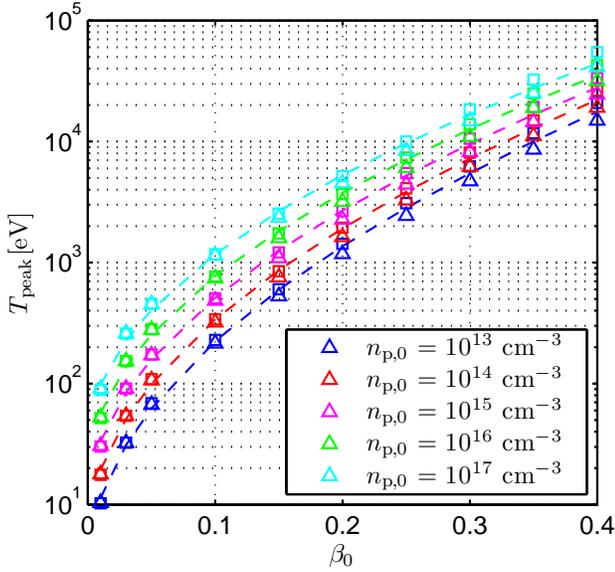}
\caption{Peak surface temperature as a function of breakout shock velocity $\bt_0$ for different values of pre-breakout ion number density $n_{\rm p,0}$. The calculated values for $n=3$ ($n=3/2$) are presented by triangles (squares). The fitting formula given by eq.~\eqref{eq:Tpeak_vs_bt_np_fit} is shown as dashed lines. \label{fig:Tmax_vs_bt0_np0}}
\end{figure}

At breakout the surface temperature rises sharply, with a peak width of $\sim t_0$, and then drops with time.
The peak surface temperature $\mathcal{T}_{\rm peak}$ as a function of the breakout parameters $\bt_0$ and $n_{\rm p,0}$ is shown in figure \ref{fig:Tmax_vs_bt0_np0} for $n=3$ and $n=3/2$.
Defining $\bt_0=0.1 \bt_{-1}$, the results can be fitted by the formulas
\begin{equation}\label{eq:Tpeak_vs_bt_np_fit}
\begin{split}
\log_{10}\left(\frac{\mathcal{T}_{\rm peak}}{\eV}\right)=&1+1.69 \bt_{-1}^{1/2} \\
&+\left(0.26-0.08 \bt_{-1}^{1/2}\right)\log_{10}\left(A n_{\rm p,15}\right), ~~ n=3 \ \\
\log_{10}\left(\frac{\mathcal{T}_{\rm peak}}{\eV}\right)=&0.95+1.78 \bt_{-1}^{1/2} \\
&+\left(0.26-0.08 \bt_{-1}^{1/2}\right)\log_{10}\left(A n_{\rm p,15}\right), ~~ n=3/2
\end{split}
\end{equation}
with an error of less than 10\% in $\mathcal{T}_{\rm peak}$ in the range $0.4>\bt_0>0.01$ and $10^{17}A^{-1}~\cm^{-3}>n_{\rm p,0}>10^{13}A^{-1}~\cm^{-3}$.
The results of the fitting formulas averaged between $n=3$ and $n=3/2$ are also shown in figure \ref{fig:Tmax_vs_bt0_np0} in dashed lines.
The averaged fitting formula is given by
\begin{equation}\label{eq:Tpeak_vs_bt_np_fit_averaged}
\begin{split}
\log_{10}\left(\frac{\mathcal{T}_{\rm peak}}{\eV}\right)=&0.975+1.735 \bt_{-1}^{1/2} \\
&+\left(0.26-0.08 \bt_{-1}^{1/2}\right)\log_{10}\left(A n_{\rm p,15}\right).
\end{split}
\end{equation}

The peak temperature time is roughly $t_{\rm peak}=0-1 t_0$ prior to the peak luminosity time, which is $t_{\rm ref}/t_0=-1.25~(-0.78)$ for $n=3~(n=3/2)$ (see Paper I).
The peak time can be described as a function of the parameter $\chi=\bt_0 (A n_{{\rm p}_0}/10^{15}~\cm^{-3})^{-1/36}$ and fitted by the formulas
\begin{equation}\label{eq:tpeak_vs_x_fit}
\begin{split}
t_{\rm peak}=&t_{\rm ref}-\left(0.83-0.85{\rm e}^{-14.1 \chi}\right)t_0, ~~ n=3 \ \\
t_{\rm peak}=&t_{\rm ref}-\left(0.7-0.8{\rm e}^{-17.75 \chi}\right)t_0, ~~ n=3/2
\end{split}
\end{equation}
with an error of less than 10\% in $t_{\rm peak}$ in the range $0.3>\chi>0.05$.

\begin{figure}[h]
\includegraphics[scale=0.8]{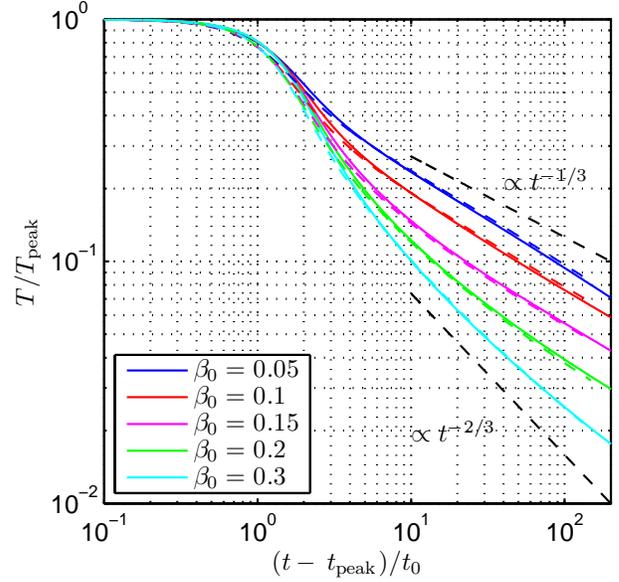}
\caption{Surface temperature as a function of time, from calculations with different values of breakout shock velocity $\bt_0$ and ion number density of $n_{{\rm p},0}=10^{15}A^{-1}~\cm^{-3}$. Calculations with $n=3$ ($n=3/2$) are shown in solid (dashed) lines. Two dashed black lines show the bounding power-laws of the temperature decline, $t^{-1/3}$ in LTE and $t^{-2/3}$ according to \cite{Nakar10}. \label{fig:T_vs_t}}
\end{figure}

Figure \ref{fig:T_vs_t} shows the surface temperature normalized to the peak value as a function of normalized time for several calculations with different values of $v_0$ and $n_{{\rm p}_0}=10^{15}A^{-1}~\cm^{-3}$.
The asymptotic behavior of the surface temperature is shown to be bounded between a power-law slope of $\propto t^{-1/3}$ and a power-law slope of $\propto t^{-2/3}$.
The temperature curves are also provided in table \ref{tab:SurfTemp} (\ref{tab:SurfTemp_n1_5}) for $n=3$ ($n=3/2$).
The derived spectral luminosity is a robust prediction of the shock breakout model, since both the temperature and the bolometric luminosity are not sensitive to the envelope density structure.

A $T\propto t^{-1/3}$ asymptotic decline is expected for radiation in thermal equilibrium or in adiabatic expansion, in agreement with the power-law upper bound presented in figure \ref{fig:T_vs_t}.
A $T\propto t^{-2/3}$ asymptotic decline was predicted by \citet{Nakar10} for the out-of-equilibrium case, in agreement with the exact solution presented here.
We note however that at late times, the production of additional photons turns out to be insignificant as opposed to the assumptions leading to the derivation of $T\propto t^{-2/3}$ in \citet{Nakar10}.

Different combinations of $v_0$ and $\rho_0$, which give the same value of $\chi$, yield approximately the same temporal dependence of the surface temperature (normalized to $\mathcal{T}_{\rm peak}$, see \sref{sec:Dimensionless parameters} for discussion). The time dependence of the surface temperature is given in tables \ref{tab:SurfTemp} and \ref{tab:SurfTemp_n1_5} for various values of $\chi$.  In order to calculate the normalized surface temperature at different times for values of $\chi$ not provided in the tables, an interpolation in $\chi$ is required.
A linear interpolation,
\begin{equation}\label{eq:T_t_interpolate}
\frac{\mathcal{T}(t,\chi)}{\mathcal{T}_{\rm peak}(\chi)}=\frac{\mathcal{T}(t,\chi_1)}{\mathcal{T}_{\rm peak}(\chi_1)}+\left(\frac{\mathcal{T}(t,\chi_2)}{\mathcal{T}_{\rm peak}(\chi_2)}-\frac{\mathcal{T}(t,\chi_1)}{\mathcal{T}_{\rm peak}(\chi_1)}\right)\frac{\chi-\chi_1}{\chi_2-\chi_1}
\end{equation}
where $\chi_2>\chi_1$ are adjacent values provided in the tables and $\chi_2>\chi>\chi_1$, yields an error of less than $5\%$ in the normalized temperature.
Eq.~\eqref{eq:Tpeak_vs_bt_np_fit} can then be used to calculate the surface temperature in physical units.
The time in physical units is obtained by using eq.~\eqref{eq:A0} and the peak surface temperature time which is approximately given by eq.~\eqref{eq:tpeak_vs_x_fit}.

\section{Emission spectrum}\label{sec:EmisSpec}

Photons diffusing outwards from an optical depth of $\tau=\beta_0^{-1}$ are found in local Compton equilibrium and are characterized by a Wien spectrum close to the peak energy.
The radiation spectrum is not significantly modified while photons diffuse to the surface, as the photons outnumber the electrons.
Consequently, the electrons' temperature equilibrates with that of the radiation without introducing a significant change to the radiation's Wien spectrum.
Therefore, the escaping photons' spectrum can be estimated from the bolometric emitted flux and the surface temperature, assuming a Wien spectrum.
Note that no high energy power law tail is expected in the radiation spectrum, as the radiation temperature is higher than or similar to the matter temperature.

The instantaneous emitted energy flux per logarithmic unit photon energy is given by
\begin{equation}\label{eq:LuminositySpectrum}
\mathcal{L}_\nu(t)=\frac{1}{6}\mathcal{L}(t)x^4 \exp(-x),~~~x=\nu/\mathcal{T}(t)
\end{equation}
where the pre-factor $1/6$ is the normalization factor, so that $\int_{-\infty}^{\infty}\mathcal{L}_\nu(t)d{\rm ln}\nu=\mathcal{L}(t)$.
The instantaneous emitted spectral flux is used to derive the spectral fluence (time integrated spectral flux) in \sref{sec:Fluence spectrum}, and the observed luminosity in different energy bands for spherically symmetric SNe in \sref{sec:Luminosity in different energy bands}.

\subsection{Fluence spectrum}\label{sec:Fluence spectrum}
\begin{figure}[h]
\includegraphics[scale=0.8]{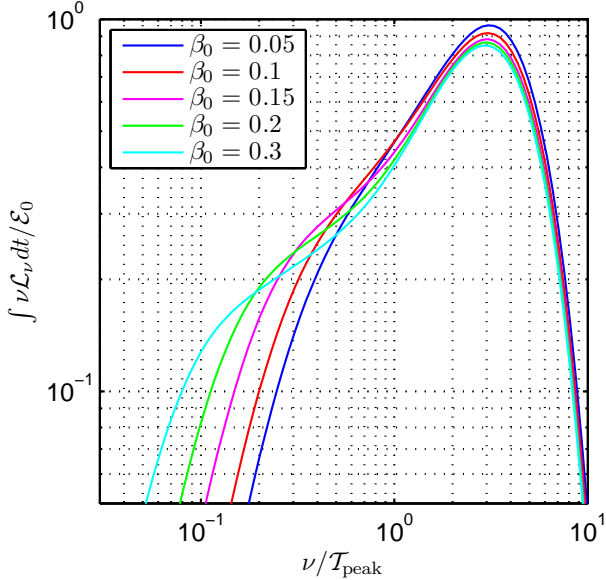}
\caption{Normalized integrated spectrum as a function of normalized photon energy, from calculations with different values of breakout shock velocity $\bt_0$, ion number density $n_{{\rm p},0}=10^{15}A^{-1}~\cm^{-3}$ and $n=3$.\label{fig:IntergatedSpecNorm}}
\end{figure}

While the observed spectrum of a SN shock breakout may be sensitive to light travel time averaging and to deviations from instantaneous arrival of the shock to the surface at all angles, a robust estimate can be obtained by considering the fluence spectrum: the flux spectrum integrated over time,
\begin{equation}\label{eq:IntegratedSpectrum}
\mathcal{E}_{\nu}(t)=\int_{-\infty}^t \mathcal{L}_\nu(t')dt'.
\end{equation}
The flux spectrum integrated up to time $(t-t_{\rm ref})/t_0=100$ for different values of $\rho_0$ and $v_0$ are shown in figure \ref{fig:IntergatedSpecNorm}. Normalizing the emitted fluence by
\begin{equation}\label{eq:E0}
\mathcal{E}_0=\kappa^{-1} \bt_0 c^2
\end{equation}
and the photon energies by the peak surface temperature, an almost universal shape is obtained for the fluence spectrum.
Therefore, the photon energy of the integrated flux peak is a good estimate for the maximal surface temperature,
\begin{equation}\label{eq:nupeak_Tpeak}
\nu_{\rm peak}=3\mathcal{T}_{\rm peak}.
\end{equation}
A relation between shock velocity and density at breakout can then be inferred from the observed fluence peak energy and eqs.~\eqref{eq:Tpeak_vs_bt_np_fit} and \eqref{eq:nupeak_Tpeak}.

The low energy part of the integrated spectrum can also be estimated from eqs. \eqref{eq:LuminositySpectrum} and \eqref{eq:IntegratedSpectrum}.
Taking $\mathcal{T}\propto t^{-\delta}$ and $\mathcal{L}\propto t^{-4/3}$, the low energy tail of the integrated spectrum has a non-thermal part with a photon energy dependence of $\mathcal{E_\nu}\propto\nu^{1/(3\delta)}$. This result can also be derived by considering $\mathcal{E}_\nu\propto \mathcal{L}_\nu t$ and taking $\mathcal{T}\approx \nu$ \citep{Nakar10}.
Thermal (Rayleigh-Jeans) contribution to the low energy tail of the instantaneous flux spectrum, with a $\nu^3$ dependence, would not change the integrated spectrum significantly, assuming $1/3<\delta<2/3$ (see \sref{sec:SurfTemp}), although a more moderate spectral slope may change this result.

We point out that the fluence spectrum is different from the instantaneous emitted flux (Wien) spectrum as well as from a black-body spectrum.
In particular, while a Wien spectrum peaks at $\nu_{\rm peak}=4\mathcal{T}_{\rm peak}$ and a black-body spectrum peaks at $\nu_{\rm peak}=2.82\mathcal{T}_{\rm peak}$, the fluence peaks at $\nu_{\rm peak}=3\mathcal{T}_{\rm peak}$.
In addition, the low energy tail at energies $\nu<3\mathcal{T}_{\rm peak}$ has a shallower slope than either Wien or black-body spectra.

\subsection{Luminosity in different energy bands}\label{sec:Luminosity in different energy bands}
Considering shock breakout from a spherically symmetric SN and taking into account light travel time integration, the observed flux in different energy bands can be derived following the derivation of Paper II for the bolometric luminosity,
\begin{equation}\label{eq:Bands}
\frac{L_{\nu,{\rm obs}}(t)}{4\pi R^2}=\int_{\nu_1}^{\nu_2}\int_0^{1} \mu h_\nu(\mu)\mathcal{L}_{\nu}\left[t-\frac{R}{c}(1-\mu)\right] F_{\rm abs} d\mu d{\rm ln}\nu,
\end{equation}
where $R$ is the stellar radius, $h(\mu)$ is the angular distribution of radiation intensity at the surface,
$\nu_1$ and $\nu_2$ are the energy band boundaries and $F_{\rm abs}(\tau_{\nu, \rm abs})$ is the absorption fraction, where $\tau_{\nu, \rm A}$ is the absorption optical depth on the line of sight as a function of photon energy.

Assuming that the radiation angular distribution is not frequency dependent and ignoring absorption, the integral over photon energies can be solved separately,
\begin{equation}\label{eq:BandsCalc}
L_{\nu,{\rm obs}}(t)=\frac{2}{3}\pi R^2\int_0^{1}\mathcal{L}(t')\mathcal{F}_\nu\left[\nu_1,\nu_2,\mathcal{T}(t')\right]h(\mu)\mu d\mu,
\end{equation}
where $t'=t-R(1-\mu)/c$, and the integral $\mathcal{F}=\int_{x_1}^{x_2}x^4{\rm exp}(-x)d{\rm ln}x$ is given by
\begin{equation}\label{eq:SpecBandsFunc}
\mathcal{F}_\nu(\nu_1,\nu_2,\mathcal{T})=\left.e^{-x}\left(x^3+3x^2+6x+6\right)\right|_{x_2=\nu_2/\mathcal{T}}^{x_1=\nu_1/\mathcal{T}}.
\end{equation}
Therefore, whenever the surface temperature crosses inside/outside an energy band boundary, the luminosity in this band lights up/drops exponentially.

\section{Shock breakout searches}\label{sec:Shock breakout searches}

Assuming all SNe are accompanied by a shock breakout, detecting the shock breakout transients and their signature can be used to reveal the properties of SN progenitors.
In order to plan a search strategy for shock breakout transients it is necessary to estimate the expected luminosity and photons' energy. Based on the exact results presented here and in Papers I-II, these values are estimated for possible SN progenitors in \sref{sec:Possible progenitors}, followed by estimates for the detection rate of shock breakout events for a number of X-ray telescopes in \sref{sec:Detection of SN breakouts}.

\subsection{Possible progenitors}\label{sec:Possible progenitors}

We focus next on red supergiants (RSGs), blue supergiants (BSGs) and He-rich Wolf-Rayet (WR) stars as possible progenitors of core collapse SNe.
RSGs and He-rich WRs have been associated with SNe of type II-p and type Ib respectively \citep{Smartt09}, while the progenitor of SN1987A has been identified as a BSG with a mass of $\sim 20 M_{\odot}$ and radius of $\sim 3\times 10^{12}~\cm$ \citep{Arnett89}.

From pre-SN images, the mass range of progenitors for type II-p SNe is constrained to $\sim 10-20 M_{\odot}$, while the progenitors' radii are $R\sim 3-7 \times 10^{13}~\cm$ \citep{Smartt04,Maund05,Mattila08,Van Dyk12}.
Using an analytical model to infer the radius of SNeIIp progenitors, different values were obtained by different authors, $R\approx 0.5-4\times 10^{13}~\cm$ by \citet{Hamuy03} and $R\approx 1.4-9 \times 10^{13}~\cm$ by \citet{Nadyozhin03}.
Both ranges are within the known radii distribution of RSGs \citep{van Belle09}.
Compared to the supergiants, WR are compact with $R\sim 10^{11}~\cm$ and have lost their hydrogen shell \citep{Langer89}.
In particular, helium rich WRs have been suggested as progenitors of type Ib SNe.

The physical parameters of the shock breakout problem, the breakout shock velocity and pre-shock density, can be estimated from the stellar parameters of each progenitor type, yielding (see equations 14 and 15 in Paper II)
\begin{align}\label{eq:MVParam_v}
\bt_0&= 0.05 M_{10}^{0.13} v_{*,8.5}^{1.13} R_{13}^{-0.26} \kappa_{0.4}^{0.13} f_{\rho}^{-0.09} &(\rm RSG)\cr
     &= 0.14 M_{10}^{0.16} v_{*,8.5}^{1.16} R_{12}^{-0.32} \kappa_{0.4}^{0.16} f_{\rho}^{-0.05} &(\rm BSG)\cr
     &= 0.23 M_{5}^{0.16} v_{*,8.5}^{1.16} R_{11}^{-0.32} \kappa_{0.2}^{0.16} f_{\rho}^{-0.05} &(\rm WR),
\end{align}
and
\begin{align}\label{eq:MVParam_rho}
\rho_0&= 2 \times 10^{-9} M_{10}^{0.32} v_{*,8.5}^{-0.68} R_{13}^{-1.64} \kappa_{0.4}^{-0.68} f_{\rho}^{0.45}~\gr~\cm^{-3} &(\rm RSG)\cr
      &= 7 \times 10^{-9} M_{10}^{0.13} v_{*,8.5}^{-0.87} R_{12}^{-1.26} \kappa_{0.4}^{-0.87} f_{\rho}^{0.29}~\gr~\cm^{-3} &(\rm BSG)\cr
      &= 2 \times 10^{-7} M_{5}^{0.13} v_{*,8.5}^{-0.87} R_{11}^{-1.26} \kappa_{0.2}^{-0.87} f_{\rho}^{0.29}~\gr~\cm^{-3} &(\rm WR),
\end{align}
where the envelope mass is $M_{\rm ej}=x M_{x} M_{\odot}$, the star radius is $R=10^{x}R_{x}~\cm$, the typical ejecta velocity is $v_*=10^x v_{*,x}~\cm~\se^{-1}$, and $f_\rho$ is a dimensionless parameter (see Paper II for discussion).
Note that the trend in shock velocity of the different progenitors is consistent with the trend in the photospheric velocities of the different SN types, ${\rm SNeIIp}<{\rm SN1987A}<{\rm SNeIb}$ (evaluated from observations made late in the spherical stage, after hydrogen/helium recombination).

Assuming that the stellar envelopes are composed of ionized hydrogen and helium, with mass fractions $X$ and $Y$ respectively, the opacity for each progenitor is taken as $\kappa=(1+X)/5~\cm^{2}~\gr^{-1}$.
For BSGs and RSGs (WRs), these mass fractions are taken as $X=0.7$ and $Y=0.3$ ($X=0$ and $Y=1$).
Using eqs. (A1)-(A2) in \citet{Calzavara04} and assuming that the core mass is small compared to the envelope mass, the parameter $f_\rho$ can be calculated for BSGs and WRs,
\begin{equation}\label{eq:frho}
\begin{split}
f_{\rho}=&0.072 \left(\frac{\mu}{0.62}\right)^4\left(\frac{L_*}{10^5 L_{\odot}}\right)^{-1}\left(\frac{M_{\rm ej}}{10M_{\odot}}\right)^3 \\
&\times \left(\frac{1+X}{1.7}\right)^{-1} \\
 &\times \left[1.35-0.35\left(\frac{L_*}{10^5 L_{\odot}}\right)\left(\frac{M_{\rm ej}}{10M_{\odot}}\right)^{-1}\left(\frac{1+X}{1.7}\right)\right]^{4},
\end{split}
\end{equation}
where $\mu=(2X+0.75Y)^{-1}$ is the mean molecular weight and $L_*=10^{x} L_{x} L_{\odot}$ is the stellar luminosity.
For RSGs the parameter $f_\rho$ is not well-constrained, and we take its value to be unity \citep[see also discussions in][]{Matzner99,Calzavara04}.

In addition, the typical ejecta velocity can be estimated as $v_*=(E_{\rm in}/M_{\rm ej})^{1/2}$ where $E_{\rm in}=10^{x} E_{x}~\erg$ is the explosion energy \citep{Matzner99}.
Unfortunately, this parameter can not be inferred directly from observations, but only estimated to an order of magnitude.
Given these relations, and using eqs.~\eqref{eq:Tpeak_vs_bt_np_fit} and \eqref{eq:nupeak_Tpeak}, the shock breakout flash would peak at X-ray energies, $\nu_{\rm peak}>0.5~\keV$, if the progenitor's radius is roughly (up to logarithmic corrections)
\begin{equation}\label{Rprogenitor_xray}
\begin{split}
R_{\rm X} \leq & 4.5\times 10^{12} M_{10}^{-1.67}E_{51}^{2.17} \\
& \times \left(\frac{1+X}{1.7}\right)^{0.5}~\cm ~~(\rm RSG)\\
R_{\rm X} \leq & 5.4\times 10^{12} M_{10}^{-1.78}E_{51}^{1.81}L_{5}^{0.16} \\
& \times \left[1.35-0.35L_5M_{10}^{-1}\left(\frac{1+X}{1.7}\right)\right]^{-0.63} \\
& \times \left(\frac{1+X}{1.7}\right)^{0.66}\left(\frac{2X+0.75Y}{1.62}\right)^{0.63} ~\cm ~~(\rm BSG, WR).
\end{split}
\end{equation}
As expressed by these equations, the maximal progenitor radius for producing an X-ray flash at shock breakout strongly depends on both the explosion energy and the progenitor's mass.
For instance, if SN1987A-like events are produced by BSGs with $R<5\times 10^{12}~\cm$, it is expected that they would flash at X-ray energies.
Additionally, RSGs with radius of $R=10^{13}~\cm$ and explosion energy of $E_{\rm in}=2\times 10^{51}~\erg$ produce an X-ray flash at shock breakout.
While the progenitors' stellar parameter distribution is unknown for each SN type, this raises the possibility that some SNeIIp may be accompanied by an X-ray flash.

For demonstration purposes, we consider the following stellar parameters:
\begin{align}\label{eq:ProgenitorParam}
M_{\rm ej}&=10 M_{\odot},~R=3\times 10^{13}~\cm &(\rm RSG)\cr
M_{\rm ej}&=10 M_{\odot},~R=3\times 10^{12}~\cm,~L_*=10^5 L_{\odot} &(\rm BSG)\cr
M_{\rm ej}&=10 M_{\odot},~R=1\times 10^{11}~\cm,~L_*=10^{5.3} L_{\odot} &(\rm WR),
\end{align}
and take for all progenitors an explosion energy of $E_{\rm in}=10^{51}~\erg$.
Estimating the peak breakout temperatures using eq.~\eqref{eq:Tpeak_vs_bt_np_fit} yields
\begin{align}\label{Tpeak_progenitors}
\mathcal{T}_{\rm peak}&=49.4~\eV &(\rm RSG)\cr
\mathcal{T}_{\rm peak}&=258~\eV &(\rm BSG)\cr
\mathcal{T}_{\rm peak}&=3.7~\keV &(\rm WR).
\end{align}
For comparison, \citet{Nakar10} predict breakout temperatures of
$T_{\rm RSG}=63.5 M_{10}^{-0.3}E_{51}^{0.5}R_{13}^{-0.65}~\eV$, $T_{\rm BSG}=1 M_{10}^{-1.2}E_{51}^{1.7}R_{12}^{-1.1}~\keV$ and $T_{\rm WR}=25.8 M_{10}^{-1.7}E_{51}^{1.8}R_{11}^{-1.5}~\keV$, yielding $T=31.1$~eV, 301~eV, and 25.8~keV for the RSG, BSG and WR progenitors we are considering.
Coincidentally, the peak temperature for the BSG progenitor in our model and in the model by \citet{Nakar10} is similar, but the dependence of the peak temperature on energy is stronger in the estimates of \citet{Nakar10} than in our results. In the explosion energy range of $E_{\rm in}=0.5-2 \times 10^{51}~\erg$ the spread in the peak temperature is a factor of 3 in our model and a factor of $\sim 10$ in the model by \citet{Nakar10}.
The differences between these results and ours lead to significant differences in the luminosities predicted in different bands and hence for outburst detectability, as discussed below.

\begin{figure}[h]
\includegraphics[scale=0.8]{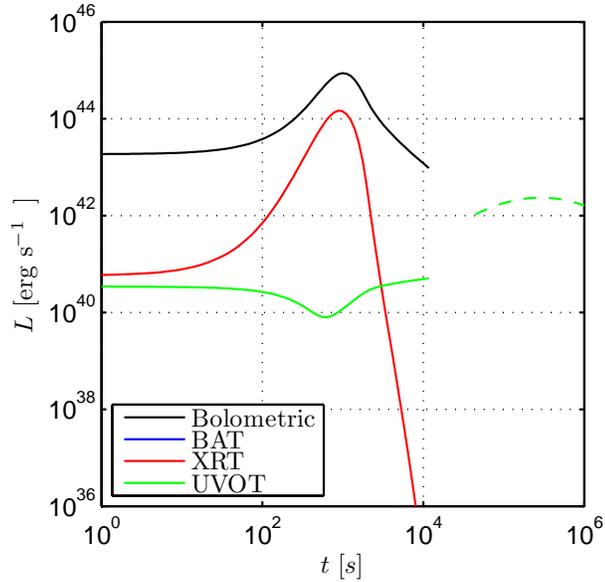}
\caption{Luminosity in different energy bands, for breakouts from an RSG associated with SNeIIp. The band luminosities in the planar stage (spherical stage, \citet{Rabinak11}) are presented in solid lines (dashed lines). Colors (black, blue, red, green) represent the different energy bands (bolometric, BAT $[15,150]$~keV, XRT $[0.3,10]$~keV, UVOT $[2,7]$~eV). For presentation purposes, the time axis is arbitrarily shifted by 1000 s with respect to $t=0$. \label{fig:BandsRSG}}
\end{figure}

\begin{figure}[h]
\includegraphics[scale=0.8]{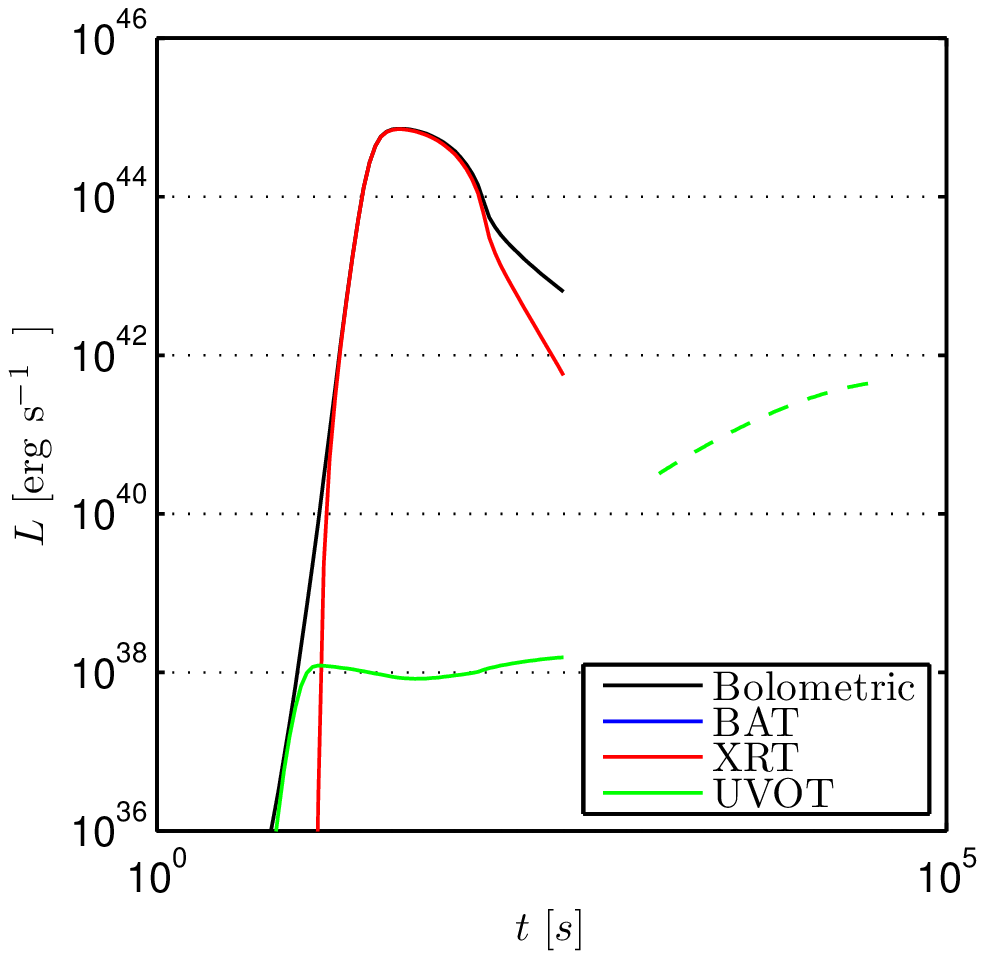}
\caption{Luminosity in different energy bands, for breakouts from a BSG associated with SN1987A. Notation is the same as figure \ref{fig:BandsRSG}. For presentation purposes, the time axis is arbitrarily shifted by 35 s with respect to $t=0$. \label{fig:BandsBSG}}
\end{figure}

\begin{figure}[h]
\includegraphics[scale=0.8]{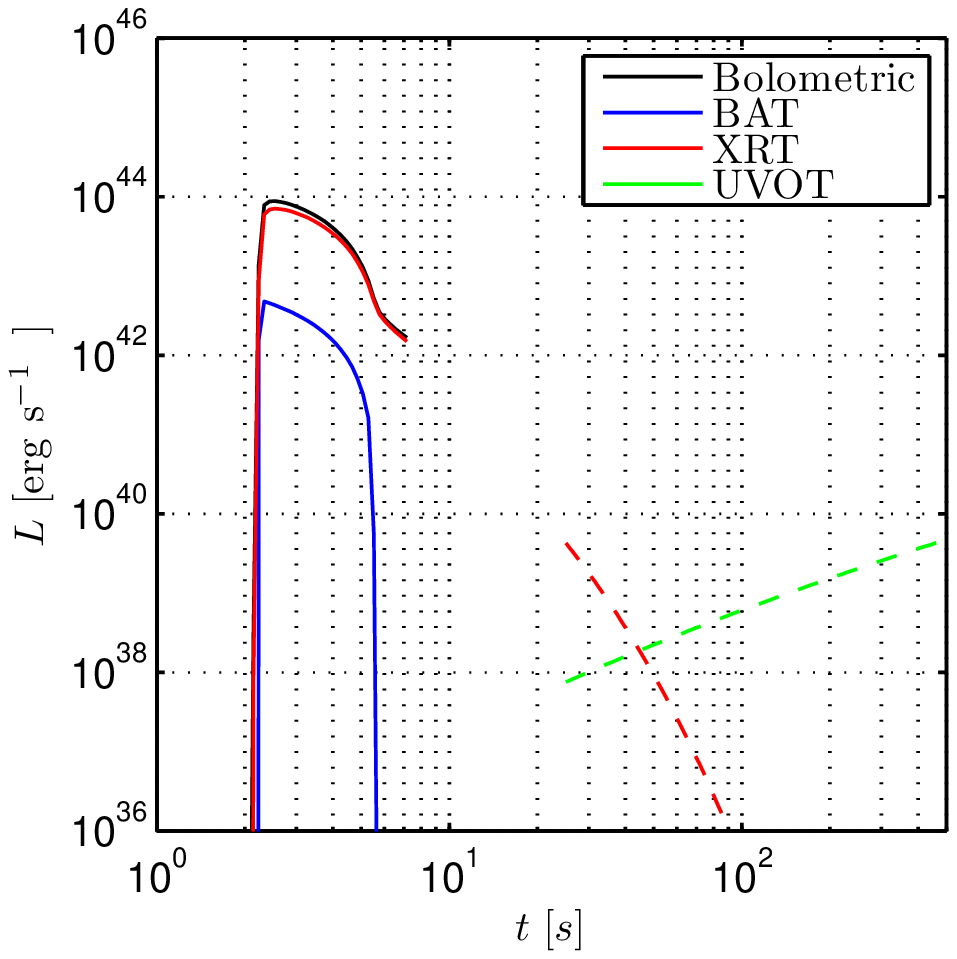}
\caption{Luminosity in different energy bands, for breakouts from a WR star associated with SNeIb. Notation is the same as figure \ref{fig:BandsRSG}. For presentation purposes, the time axis is arbitrarily shifted by 2.5 s with respect to $t=0$. \label{fig:BandsWR}}
\end{figure}

The luminosities in different energy bands following breakouts from RSGs, BSGs and WRs are presented in figures \ref{fig:BandsRSG}-\ref{fig:BandsWR} for the energy bands of the burst alert telescope (BAT) ($15~\keV<\nu<150~\keV$), the X-ray telescope (XRT) ($0.3~\keV<\nu<10~\keV$) and the optical/UV telescope (UVOT) ($2~\eV<\nu<7~\eV$) on board \textit{Swift}.
The band luminosities are calculated using eq.~\eqref{eq:BandsCalc} up to time $t_{\rm s}=R/(4v_0)$ when the planar approximation is no longer valid (see Paper II).
For completeness, the band luminosities in the spherical stage, calculated according to \citet{Rabinak11}, are also presented in the figures.

It is apparent from figures \ref{fig:BandsBSG} and \ref{fig:BandsWR} that most of the prompt emission (on a timescale $R/c$) in BSGs and WRs is in the XRT band.
Figure \ref{fig:BandsWR} also shows that for WRs there is also a smaller contribution to the emission by the BAT band.
Note though that this result is sensitive to the assumed explosion energy, as higher values of this parameter imply that the emission would be dominated by the BAT band.
In RSGs, although the emission following breakout peaks below the lower energy of the band, $\nu_{\rm peak}=3 \mathcal{T}_{\rm peak}<0.15~\keV$, there is still significant emission in the XRT band (see figure \ref{fig:BandsRSG}).
This is a result of eq.~\eqref{eq:SpecBandsFunc}, as the luminosity radiated in the XRT band is a fraction $(x^3+3x^2+6x+6){\rm e}^{-x}/6$ of the bolometric luminosity, where $x=\nu_1/\mathcal{T}(t)$ and $\nu_1$ is the lower energy of the band.

Essentially, the total number of photons emitted in the burst determines the detectability of the transient.
Approximating the emitted number of photons as $N_\gamma = 4 \pi R^2 \mathcal{E}_0 / \mathcal{T}_{\rm peak}$, yields for the different progenitors
\begin{align}\label{eq:Ngamma_progenitors}
N_\gamma&=9\times 10^{57} &(\rm RSG)\cr
N_\gamma&=5.2\times 10^{55} &(\rm BSG)\cr
N_\gamma&=1.6\times 10^{52} &(\rm WR).
\end{align}
This approximation is good to better than $5\%$ accuracy for an energy band between $\mathcal{T}_{\rm peak}/3<\nu<5\mathcal{T}_{\rm peak}$.
Note that the larger the radius of the progenitor, the lower is the breakout shock velocity (see \eqref{eq:MVParam_v}) but the higher is the energy released.
Consequently, as the peak temperature is also lower, the total number of emitted photons is higher, which makes the event easier to detect.
However, as the peak temperature is lower, the emitted photons may be absorbed by the interstellar medium.
Considering a neutral hydrogen column density of $N_{\rm H}=1 \times 10^{21}~\cm^{-2}$, the optical depth for photoionization is $>1$ for $\nu<0.5~\keV$ \citep{Wilms00}, and photons emitted at lower energies are efficiently absorbed.

Even though the emission following breakout from an RSG with radius $R=3\times 10^{13}~\cm$ peaks at energy $\nu_{\rm peak}=3 \mathcal{T}_{\rm peak}<0.15~\keV$, the number of emitted photons is so high that it is still observable in the XRT band if $N_{\rm H}\leq 1 \times 10^{21}~\cm^{-2}$.
This is also seen in figure \ref{fig:BandsRSG}.
Using eqs.~\eqref{eq:IntegratedSpectrum} and \eqref{eq:BandsCalc}, the number of observed photons is diminished by a factor $(x^2+2x+2){\rm e}^{-x}/2$ compared to the number emitted, where $x=\nu_1/\mathcal{T}_{\rm peak}$ and $\nu_1$ is the lower energy of the band.
However, in this case only the high energy tail of the fluence spectrum would be observable, making a definite identification of the transient as a shock breakout event difficult.

Thus, the best prospects for detecting a shock breakout flash are from BSG progenitors with radii $R\approx 3\times 10^{12}~\cm$, the burst from which is characterized by a relatively long duration, $\sim R/c\lesssim 100~\se$, and peaks in low X-ray energies, $\nu_{\rm peak} =3\mathcal{T}_{\rm peak} \approx 0.8~\keV$.
In addition, type II-p SNe may be accompanied by an observable X-ray flash if the neutral hydrogen column density is $N_{\rm H}<1 \times 10^{21}~\cm^{-2}$ and the progenitor's radius is $R<10^{13}~\cm$ as proposed for some SNeIIp by \citet{Hamuy03}.

\subsection{Detection of SN breakouts}\label{sec:Detection of SN breakouts}

In general, detection of a transient is determined by the background and by the sensitivity of the detector.
The X-ray background in the energy band $0.3~\keV \leq \nu \leq 10~\keV$ is $n_{\rm CXB} \approx 0.01~{\rm counts}~\cm^{-2}~{\rm deg}^{-2}~\se^{-1}$ \citep{De Luca04,Hickox06}, while in the energy band $15~\keV \leq \nu \leq 100~\keV$ it is $n_{\rm CXB} \approx 7.5 \times 10^{-4}~{\rm counts}~\cm^{-2}~{\rm deg}^{-2}~\se^{-1}$ \citep{Ajello08}.
The number of photon counts $n_{\rm det}$ needed for a significant detection of a transient with duration $\Delta t$ with a detector with an effective area $A_{\rm eff}$, PSF angular size $\Phi$, field of view $\Omega$ and lifetime $t_{\rm I}$ is determined by Poisson statistics, $(a^{n_{\rm det}}{\rm e}^{-a})/n_{\rm det}!=L^{-1}$,
where $a=n_{\rm CXB} (A_{\rm eff}/\cm^2)(\Phi/{\rm deg}^2)(\Delta t/\se)$ and $L=(t_{\rm I}/\Delta t)(\Omega/\Phi)$.
Because of the low X-ray background, $n_{\rm det}<10-25$ for active and non-active X-ray telescopes (see table \ref{tab:X-ray telescopes} for X-ray telescopes characteristics).
However, $n_{\rm det}$ only represents a lower limit on the number of photon counts needed for identifying a transient over exposure of duration $\Delta t$.
To identify the transient with SN shock breakout it is necessary also to compare the total emitted energy and the fluence spectrum of the event with the breakout model.
We therefore arbitrarily set the lower limit on the number of detected photons at $n_{\rm det}=100$ in order to determine the nature of the event according to the fluence spectrum.

Assuming emission of $N_\gamma$ photons in a stated energy band, such events are detectable to a distance of
\begin{equation}
d_{\rm det}= 578 \left(\frac{N_\gamma}{10^{55}}\right)^{1/2}\left(\frac{A_{\rm eff}}{400~\cm^2}\right)^{1/2}\left(\frac{n_{\rm det}}{100}\right)^{-1/2}~\Mpc.
\end{equation}
Taking the shock breakout rate per unit volume to be $\dot{n}_{\rm B}$, the number of shock breakout events detected is
\begin{equation}\label{eq:Nbreakouts}
\begin{split}
N_{\rm SNSB} =& \frac{\Omega}{3} d_{\rm det}^3 \dot{n}_{\rm B} t_{\rm I} \\
=& 0.2 \left(\frac{\Omega}{{\rm deg}^2}\right)\left(\frac{N_\gamma}{10^{55}}\right)^{3/2}\left(\frac{A_{\rm eff}}{400~\cm^2}\right)^{3/2}  \\
& \times \left(\frac{n_{\rm det}}{100}\right)^{-3/2} \left(\frac{\dot{n}_{\rm B}}{10^{-5}~\Mpc^{-3}~\yr^{-1}}\right) \left(\frac{t_{\rm I}}{\yr}\right).
\end{split}
\end{equation}
Table \ref{tab:X-ray telescopes} presents the expected number of shock breakout detections over the lifetime of several active and non-active X-ray telescopes for SN1987A-like events, taking $n_{\rm det}=100$ and different explosion energies.
The shock breakout rate is calculated from the total core collapse SN rate per unit volume, $\sim 3\times 10^{-4}~\Mpc^{-3}~\yr^{-1}$ \citep{Dahlen12}, and the SN1987A-like events fraction of that rate, which is $\sim 3\%$ \citep{Smartt09}.

Although SNeIIp comprise about 59\% of the core collapse SN rate \citep{Smartt09}, their detection rate is not reported here as their breakout emission is expected to be absorbed by the interstellar medium.
In a similar fashion, SNeIb comprise 10\% of the core collapse SN rate, but the number of photons emitted at breakout is too low to detect.
If the progenitors of this SN type are smaller than what was assumed here, or the explosion energy is higher, than the emitted burst would peak at higher energies and might be detected by the BAT on-board \textit{Swift}.

As can be seen from table \ref{tab:X-ray telescopes}, \textit{XMM-Newton} and \textit{ROSAT} are the best candidates for detecting SN shock breakouts from SN1987A-like events, with $>30$ events over the lifetime of each telescope.
This means that there are expected to be many shock breakout events in the archival data of each of these telescopes (\citet{Calzavara04} come to a similar conclusion based on breakout calculations by \citet{Blinnikov00}).
We note however that previous transient surveys conducted with the archival data of \textit{ROSAT} did not detect such bursts \citep{Vikhlinin98,Greiner00}.
This poses a problem for the stellar envelope breakout scenario, as the predictions of total emitted energy and total emitted photon number are a robust result of the model, and mainly depend on the typical ejecta velocity and the progenitor's radius.
Furthermore, both the bolometric and temperature results are not sensitive to the density structure.
Lack of shock breakout detections may indicate that either the typical parameters assumed for the progenitors are grossly incorrect, or that the progenitors are not characterized by a hydrostatic stellar envelope at the end of their lives (with a density profile monotonically decreasing towards the surface).

\section{Previous works and observations}\label{sec:Previous works and observations}

We provide here a comparison to previous works in the literature describing the temperature expected in shock breakouts from stellar envelopes, as well as a comparison to the proposed shock breakout candidate XRF080109.

\subsection{Temperature calculations in the literature}\label{sec:Temperature calculations in the literature}
Numerical works in the past have provided solutions for the emission following shock breakout from specific progenitor profiles, but mostly for explosions with low ejecta velocity, $\bt_0 \lesssim 0.05$ \citep{Ensman92,Blinnikov00}.
For example, \citet{Blinnikov00} have calculated the expected breakout temperature for models describing the SN1987A progenitor, using pre-supernova density profiles which are based on stellar evolution models.
However, these calculations did not incorporate incoherent Compton scattering.

We compare here our results with the ones reported for a BSG progenitor model with $\{M=14.67 M_{\odot}, R=48.5 R_{\odot} \}$.
For $E=(0.72, 1.02, 1.34) \times 10^{51}~\erg$, \citet{Blinnikov00} report a maximum bolometric luminosity of $L_{\rm max}= (4.2, 6.8, 9.5) \times 10^{44}~\erg~\se^{-1}$, a maximum ejecta velocity of $v_{\rm max}=(2.8, 3.3, 3.7) \times 10^9 ~\cm~\se^{-1}$ and a maximum temperature of $T \approx (90 ,105, 115)~\eV$, which roughly scales as $L_{\rm max}^{1/4}$.
Our theoretical model for shock breakout from a gas with a power-law density profile and constat opacity provides similar results for the bolometric properties, but different ones for the temperature.
Using the same set of parameters and taking $v_*=(E/M)^{1/2}, \kappa=0.34$ and $f_{\rho}=0.02$, the maximum bolometric luminosity and maximum ejecta velocity are $(4.3, 6.8, 9.6) \times 10^{44}~\erg~\se^{-1}$ and $(2.5, 3.1, 3.6) \times 10^9~\cm~\se^{-1}$, respectively (see table 2 in Paper I and eqs. 7 \& 25 in Paper II).
Contrary to the bolometric properties, the difference in temperature is significant: according to eq.~\eqref{eq:Tpeak_vs_bt_np_fit}, the maximum temperature is predicted to be $T \approx (153, 202, 254)~\eV$.

Different predictions are obtained from the analysis of \citet{Nakar10}, which provides analytical estimates for the breakout temperature, taking into account Comptonization of the bremsstrahlung emitted photons.
However, their calculation assumes only local photon production, without accounting for diffusion or expansion.
Using eq. 38 therein and for the same sets of parameters, the radiation is (marginally) out of thermal equilibrium, and the peak temperature is $T \approx (95, 175, 274)~\eV$.
Therefore, our results are closer to the ones given by \citet{Nakar10}, but the spread in maximum temperature between different sets of parameters according to \citet{Nakar10} (\citet{Blinnikov00}) is higher (lower) than our results which include incoherent scattering, photon diffusion and expansion.

\subsection{SN2008D as a shock breakout candidate}\label{sec:SN2008D as a shock breakout candidate}
X-ray flash (XRF) 080109/SN2008D has been proposed as a shock breakout candidate \citep{Soderberg08,Chevalier08,Katz10}, mainly based on its energetics (total released energy, typical luminosity).
This XRF is characterized by a peak X-ray luminosity of $10^{43}~\erg~\se^{-1}$ in the $0.3-10~\keV$ band and $\sim 300~\se$ duration, and a total emitted energy of $E \sim 2.5\times 10^{45}~\erg$.

The observed XRF fluence spectrum can be fitted by the universal fluence function (see eq.~\eqref{eq:IntegratedSpectrum} and figure \ref{fig:IntergatedSpecNorm}), with $\nu_{\rm peak}=3.9~\keV$, and is consistent with the breakout fluence spectrum up to photon energy of $\sim 8~\keV$.
The fit is somewhat in conflict with detection of photons above this energy, albeit with a small number of counts.
In addition, an almost flat photon energy spectrum, $\mathcal{E}_{\nu} \sim \nu^{0}$ was suggested to agree with the observations \citep{Soderberg08,Modjaz09}, though it also has some tension with photons detected at energies $>6~\keV$. If the emitted spectrum is indeed flat, then this event is not consistent with a shock breakout from a stellar envelope interpretation. Even without assuming a Wien spectrum for the emission, the expected low energy tail of the fluence should be characterized by a positive spectral slope, as both the luminosity and the radiation temperature drop with time (see \sref{sec:Fluence spectrum}). A negative or flat slope would indicate that the emitted photons are not characterized by a specific radiation temperature, or that the main emission mechanism is accompanied by a high energy power-law tail, both explanations are in disagreement with the stellar envelope shock breakout scenario.

Assuming that the observed spectrum may be interpreted as a breakout spectrum, ignoring the above mentioned tensions between the model and the observations, $\nu_{\rm peak}=3.9~\keV$ implies, using eqs.~\eqref{eq:nupeak_Tpeak} and \eqref{eq:Tpeak_vs_bt_np_fit}, $\bt_0 \approx 0.12-0.17$ (the variation is due to the uncertainty in $n_{{\rm p},0}$).
This implies a terminal ejecta velocity of $v_{\rm ej}=(0.24-0.34) c$, which agrees with radio observations \citep{Soderberg08}. Assuming $\kappa=0.2$, the progenitor's radius can then be found using the total emitted energy (see eq. 40 in Paper II)
\begin{equation}
R=\left(\frac{1}{8\pi}\frac{\kappa E}{\bt_0 c^2}\right)^{1/2} \approx 5\times 10^{11}~\cm,
\end{equation}
where the correction of the emitted energy due to absorption is included and is only $\sim 50\%$. The inferred radius is larger than a WR radius associated with this type Ib SN, and is also a factor 5 larger than the value inferred from UV/optical observations in the spherical stage \citep{Rabinak11}. This radius also disagrees with the observed duration of the XRF, being a factor 20 smaller than the radius related to the duration, $R_\Delta=c \Delta t \approx 10^{13}~\cm$.

If this event was indeed a shock breakout, the latter discrepancy can be explained either by assuming an non-spherical shock wave reaching different points on the surface at different times, or by assuming the stellar envelope is enshrouded by a moderately optically thick circumstellar material, with $\tau_{\rm csm} \ll \bt_0^{-1}$.
The presence of a moderately optically thick material can change the duration of the observed burst, as this is determined by the light travel time across the radius of the last scattering surface.
However, it does not significantly affect the total emitted energy or the fluence spectrum, as the extra mass is negligible, $M_{\rm csm} \ll (\kappa \bt_0)^{-1}$, and no extra energy is added by the interaction \citep{Nakar10}.

\section{Conclusions}\label{sec:Conclusions}
Solutions for the temperature and photon number density were calculated for the problem of an RMS breaking out of a gas with a power law density profile, assuming constant Thomson scattering opacity, local Compton equilibrium and bremsstrahlung as the dominant emission rate.
The radiation temperature is determined by three dimensional parameters: breakout shock velocity $v_0$, (pre-shock) density $\rho_0$ and opacity $\kappa$.
Calculations presented here are the same for envelopes with any hydrogen-helium composition, as the results depend on the atomic fractions of the elements $x_i$ and their atomic numbers $Z_i$ and weights $A_i$ only through the combination $\Sigma x_i Z_i^2/\Sigma x_i A_i$.

The temperature at the surface, at an optical depth $\tau v_0/c=0$, is similar to the temperature at an optical depth $\tau=c/v_0$ (see figure \ref{fig:T_vs_M_2}).
The surface temperature rises with time to a sharp peak of width $\sim t_0=c/(\rho_0 v_0^2 \kappa)$, at a time $(0.1 - 1) t_0$ prior to maximum bolometric luminosity.
The relation between peak surface temperature, breakout shock velocity and density is well approximated by eq.~\eqref{eq:Tpeak_vs_bt_np_fit}.
Tables \ref{tab:SurfTemp} and \ref{tab:SurfTemp_n1_5} tabulate the surface temperature as a function of time for different values of the parameter $\chi=(v_0/c)(\rho_0/(m_{\rm p})/10^{15}~\cm^{-3})^{-1/36}$.
These tables can be used in order to calculate the temperature for different $v_0$ and $\rho_0$ values not provided here, using linear interpolation in the parameter $\chi$ (see eq.~\eqref{eq:T_t_interpolate}).

Assuming a Wien spectrum of the escaping photons, the instantaneous emission spectrum can be calculated using the bolometric luminosity and the surface temperature (see eq.~\eqref{eq:LuminositySpectrum}).
Accordingly, the luminosity in different energy bands can be calculated using eq.~\eqref{eq:Bands}.
The spectral fluence, the integral over time of the spectral flux, shows a universal behavior (see figure \ref{fig:IntergatedSpecNorm}), up to scaling of the fluence by $\mathcal{E}_0=\kappa^{-1} v_0 c$ and of the photon energy by the peak surface temperature $\mathcal{T}_{\rm peak}$.
The predicted breakout fluence spectrum is non-thermal, characterized by a positive spectral slope in the low energy tail, and a peak energy at $\nu_{\rm peak}\approx 3\mathcal{T}_{\rm peak}$.

The light curve in different energy bands for typical progenitors, corresponding to suggested progenitors of different SN types, was presented in figures \ref{fig:BandsRSG}-\ref{fig:BandsWR}. It was shown that the prompt emission (on timescale $R/c$) following breakout from BSGs is mostly in the XRT band ($0.3-10~\keV$), while for WRs it starts in the BAT band ($15-150~\keV$) and then falls into the XRT band at times $t\approx R/(4v_0)$.
RSGs have a peak temperature lower than the energy band, but radiate significantly also in the XRT band ($L\sim 10^{44}~\erg~\se^{-1})$. Our results differ from those of \citet{Nakar10}, with a growing discrepancy as the radiation is driven further away from equilibrium. In particular, the results of \citet{Nakar10} indicate that it would be difficult to observe RSG \& WR breakouts in the XRT band, as their predicted temperatures for RSG and WR progenitors with parameters similar to those used here are closer to the UV and BAT bands respectively.

We have compared our results to the numerical simulations of \citet{Blinnikov00}, and to the corresponding analytical estimates of \citet{Nakar10}, for a given pre-supernova density profile assumed to describe the 1987A progenitor.
The peak bolometric luminosities and maximum ejecta velocities obtained in the numerical simulations are consistent with our results. The peak temperature in the numerical simulations, which do not include incoherent Compton scattering, roughly scales with maximum luminosity as $L_{\rm max}^{1/4}$, while the peak temperature in the analytical work of \citet{Nakar10}, which takes into account the Comptonization of photons but not photon diffusion or the effect of hydrodynamic expansion, offers a bigger spread as a function of maximum luminosity. The results of this work are closer to the analytic estimates, but provide a smaller (larger) spread in the peak temperature as a function of maximum luminosity than the analytical (numerical) results.

The detection rate of SN breakouts was evaluated for SN1987A-like events, considering the characteristics of different X-ray telescopes (see table \ref{tab:X-ray telescopes} calculated using eq.~\eqref{eq:Nbreakouts}).
$>30$ SN1987A-like breakout detections are expected over the lifetimes of \textit{ROSAT} and \textit{XMM-Newton}.
If such events are indeed not detected it would indicate that the typical parameters assumed for the BSG progenitors are grossly incorrect, or that the progenitors' density profile does not monotonically decrease towards the surface, as expected for a hydrostatic stellar envelope.

An analysis of the SN2008D/XRF 080109 event was performed, indicating that the observed fluence spectrum may be consistent with the breakout interpretation, provided that photons detected at energies $>8~\keV$ are a statistical coincidence.
Inferring the shock velocity from the peak fluence energy according to the universal breakout fluence, implies a terminal ejecta velocity which agrees with radio observations \citep{Soderberg08}.
On the other hand, the inferred radius of the event is $R \approx 5\times 10^{11}~\cm$, a value much higher than the expected radii of SNeIb progenitors and also higher than the value inferred from optical/UV observations in the spherical stage \citep{Rabinak11}, and much lower than the value suggested by the XRF duration.

\acknowledgments
We thank David Burrows, Maryam Modjaz and Alicia Soderberg for useful discussions.
N.S. and E.W are partially supported by UPBC and GIF grants.
B.K. is supported by NASA through Einstein Postdoctoral Fellowship awarded by the Chandra X-ray Center, which is operated by the Smithsonian Astrophysical Observatory for NASA under contract NAS8-03060.

\appendix

\section{A. Photon production processes}\label{sec:appendix1}
Besides bremsstrahlung emission there are more photon production processes that are relevant for a completely ionized plasma.
We show here that bremsstrahlung emission is the dominant emission processes in the temperature and density range considered ($10~\eV<T<50~\keV$, $10^{12}~\cm^{-3}<n_{\rm p}<10^{17}~\cm^{-3}$).

Bremsstrahlung emission is a two-body electron-ion radiative process with spectral emissivity (photon number density production rate per logarithmic unit photon energy) given by \citep{Rybicki86}
\begin{equation}\label{eq:ngadot_B}
\dot{n}_{\nu,\rm B}=\left(\frac{8}{3\pi}\right)^{1/2} n_{\rm p} (n_{\rm e} \sigma_{\rm T} c)\alpha_{\rm e} Z^2 \left(\frac{T}{m_{\rm e} c^2}\right)^{-1/2} \log\left(\frac{2.25T}{\nu}\right),
\end{equation}
in the limit of low photon energies $\nu \ll T$, where $n_{\rm e}$ is the electron number density.
Note that photon production by electron-electron bremsstrahlung is roughly a factor $T/m_{\rm e}c^2$ lower than by electron-ion bremsstrahlung.

Double Compton emission (or radiative Compton scattering) is a two-body electron-photon process which produces photons with an emission spectrum similar to bremsstrahlung.
Assuming the radiation is characterized by a Wien spectrum with temperature equal to the the electrons' temperature, and $T\ll m_{\rm e}c^2$, the double Compton spectral emissivity is given by \citep{Svensson84}
\begin{equation}\label{eq:ngadot_dc}
\dot{n}_{\nu,\rm DC}=\frac{16}{\pi} n_{\gamma} (n_{\rm e} \sigma_{\rm T} c)\alpha_{\rm e}\left(\frac{T}{m_{\rm e} c^2}\right)^2 g_{\rm DC}(T)
\end{equation}
where
\begin{equation}
g_{\rm DC}(T)=\left[1+13.91\frac{T}{m_{\rm e} c^2}+11.05\left(\frac{T}{m_{\rm e} c^2}\right)^2+19.92\left(\frac{T}{m_{\rm e} c^2}\right)^3\right]^{-1}.
\end{equation}
Note that the expression in eq.~\eqref{eq:ngadot_dc} is an approximation for the soft photons emission, and exhibits a logarithmic infrared divergence.
In addition, both bremsstrahlung and double Compton processes drop exponentially at $\nu\approx T$.

Considering that radiation is found at local Compton equilibrium, the local photon to ion ratio at the shock's downstream is given by
\begin{equation}
\frac{n_\gamma}{n_{\rm p}}=\frac{6}{49}A \beta_{\rm sh}^2\frac{m_{\rm p} c^2}{T}.
\end{equation}
The ratio of bremsstrahlung to double Compton emission down to the low cutoff energy $\nu_{\rm c}(n_{\rm p},T)$ (see eq.~\eqref{eq:nuc1}) is therefore given by
\begin{equation}
\frac{\dot{n}_{\rm B}}{\dot{n}_{\rm DC}} \approx 0.75 \frac{Z^2}{A} \beta_{\rm sh}^{-2} \frac{m_{\rm e}}{m_{\rm p}} \left(\frac{T}{m_{\rm e} c^2}\right)^{-3/2}\log\left(\frac{2.25T}{\nu_{\rm c}}\right) g_{\rm DC}^{-1}(T).
\end{equation}
Assuming a shock velocity of $\bt_0=0.4$, the ratio is $\dot{n}_{\rm B}/\dot{n}_{\rm DC}>10$ for $T<10~\keV$.
At the high end of the temperature and density range, $T=50~\keV$ and $n_{\rm p}=7\times 10^{17}~\cm^{-3}$, the photon production rate by bremsstrahlung emission is still 3 times higher than that of double Compton emission.
Note that for shock velocity of $\bt_0=0.4$ and power-law index $n=3/2$ a peak temperature of $50~\keV$ is reached at breakout (see figure \ref{fig:Tmax_vs_bt0_np0}), but most of the photons are produced in the downstream where the temperatures are lower.
Therefore, the temperature results can have an error of order $10\%$ at high shock velocities $\bt_0 \gtrsim 0.3$.

An additional process for photon production is radiative recombination (or free-bound emission), which is also a two-body electron-ion process.
In this case the photon production rate is determined by the recombination rate, up to a correction of order unity due to the de-excitation cascade.
Note that this radiative process does not have an infrared divergence, so the photon production rate is bound.
Assuming hydrogenic ions and a unity gaunt factor, the total photon emissivity can be approximated as \citep{Seaton59}
\begin{equation}
\dot{n}_{\rm R} = \sum_{n=1}^{\infty} \left(\frac{2}{3\pi}\right)^{1/2} \alpha_{\rm e}^{3} Z^2  n^{-3} n_{\rm p} (n_{\rm e} \sigma_{\rm T} c) \left(\frac{T}{m_{\rm e}c^2}\right)^{-3/2}{\rm e}^{Z^2 I_{\rm H}/n^2 T} E_1\left(\frac{Z^2 I_{\rm H}}{n^2 T}\right),
\end{equation}
where $I_{\rm H}=0.5\alpha_{\rm e}^2 m_{\rm e} c^2 \approx 13.6~\eV$ and $E_1(x)$ is again the exponential integral function. An important difference in the spectrum between free-bound emission and bremsstrahlung emission is that free-bound emission peaks at $\nu=T+Z^2I_{\rm H}$ and bremsstrahlung emission peaks at $\nu=T$. This means that for low temperatures, $T \lesssim Z^2 I_{\rm H}$, free-bound emission produces photons with energies higher than the typical plasma temperature. These photons then down-scatter in energy.
On the other hand, for such low temperatures the soft bremsstrahlung photons can not up-scatter in energy before being absorbed.
Thus, both bremsstrahlung and free-bound photon production rates simply drop with temperature increase at low temperatures.
However, at some turn-off point in temperature photons can up-scatter in energy, and bremsstrahlung emission starts increasing with temperature.
The exact turn-off point depends on the density, and it is between $30-300~\eV$ for the density range considered.

At $T=10~\eV$ the bremsstrahlung and the free-bound photon production rates are similar for hydrogen, and for helium the bremsstrahlung emission is 3 times higher.
As free-bound emission drops faster with temperature than bremsstrahlung emission, at the turn-off point free-bound emission becomes negligible in comparison to bremsstrahlung emission.
Moreover, free-bound emission of metals is not important for low abundance atoms.
For instance, considering completely ionized iron at $T=1~\keV$, the iron free-bound photon production rate is a factor $(3-10)n_{\rm H}/n_{\rm Fe}$ lower than the bremsstrahlung emission of hydrogen, where $n_{\rm H}/n_{\rm Fe}$ is the ratio of the hydrogen to iron ions.

\section{B. Numerical values of $\mathcal{T}(t)$}\label{sec:appendix2}
Numerical results for the surface temperature as a function of time,
$\mathcal{T}(t)$, for hydrogen-helium envelopes with $n=3$ and $n=3/2$ and several values of breakout shock velocity and pre-shock ion number density of $n_{{\rm_p},0}=10^{15}~\cm^{-3}$ are presented in tables \ref{tab:SurfTemp} and \ref{tab:SurfTemp_n1_5}. The values presented are the surface temperature normalized by the peak surface temperature (which can be fitted by eq.~\eqref{eq:Tpeak_vs_bt_np_fit}), and the time shifted by the time of peak surface temperature, normalized by $t_0$.
In order to find the time dependent temperature for different values of shock velocity and density, the results can be linearly interpolated in the parameter $\chi=\bt_0 (A n_{{\rm p},0}/10^{15}~\cm^{-3})^{-1/36}$, using eq.~\eqref{eq:T_t_interpolate}.
The time dependent temperature can then be used to determine the observed spectral luminosity (eq.~\eqref{eq:LuminositySpectrum}), fluence spectrum (eq.~\eqref{eq:IntegratedSpectrum}) and luminosity in different energy bands (eq.~\eqref{eq:Bands}).

\begin{table*}[H]
\begin{tabular}{llllll}
$(t-t_{\rm peak})/t_0$\footnote{Time relative to $t_{\rm peak}$, the time when $\mathcal{T}(t)$ peaks, normalized to $t_0=c/(\kappa\rho_0\vt_0^2)$.}&$\mathcal{T}/\mathcal{T}_{\rm peak}$\footnote{Surface temperature, normalized to peak surface temperature.}~~$(\chi=0.05)$\footnote{$\chi=\bt_0 (A n_{{\rm p},0}/10^{15}~\cm^{-3})^{-1/36}$}&$\mathcal{T}/\mathcal{T}_{\rm peak}~~(\chi=0.1)$&$\mathcal{T}/\mathcal{T}_{\rm peak}~~(\chi=0.15)$&$\mathcal{T}/\mathcal{T}_{\rm peak}~~(\chi=0.2)$&$\mathcal{T}/\mathcal{T}_{\rm peak}~~(\chi=0.3)$\\
\hline
 -1.75 & 0.0866 & 0.0838 & 0.0722 & 0.0578 & 0.0378\\
 -1.50 & 0.211 & 0.225 & 0.187 & 0.156 & 0.116\\
 -1.25 & 0.432 & 0.425 & 0.385 & 0.351 & 0.304\\
 -1.00 & 0.651 & 0.647 & 0.628 & 0.609 & 0.578\\
 -0.75 & 0.818 & 0.823 & 0.82 & 0.815 & 0.805\\
 -0.50 & 0.925 & 0.931 & 0.932 & 0.932 & 0.93\\
 -0.25 & 0.983 & 0.984 & 0.985 & 0.985 & 0.985\\
 -0.10 & 0.997 & 0.998 & 0.998 & 0.998 & 0.998\\
  0.00 & 1 & 1 & 1 & 1 & 1\\
  0.10 & 0.997 & 0.998 & 0.998 & 0.998 & 0.998\\
  0.25 & 0.984 & 0.986 & 0.987 & 0.987 & 0.987\\
  0.50 & 0.941 & 0.946 & 0.948 & 0.949 & 0.95\\
  0.75 & 0.877 & 0.883 & 0.887 & 0.888 & 0.889\\
  1.00 & 0.802 & 0.806 & 0.808 & 0.809 & 0.81\\
  1.25 & 0.725 & 0.722 & 0.721 & 0.721 & 0.719\\
  1.50 & 0.653 & 0.642 & 0.635 & 0.632 & 0.627\\
  1.75 & 0.591 & 0.571 & 0.557 & 0.55 & 0.542\\
  2.00 & 0.54 & 0.512 & 0.491 & 0.48 & 0.469\\
  2.25 & 0.498 & 0.464 & 0.437 & 0.423 & 0.409\\
  2.50 & 0.463 & 0.425 & 0.394 & 0.377 & 0.361\\
  2.75 & 0.436 & 0.394 & 0.359 & 0.34 & 0.323\\
  3.00 & 0.413 & 0.369 & 0.331 & 0.311 & 0.292\\
  3.50 & 0.377 & 0.331 & 0.289 & 0.267 & 0.247\\
  4.00 & 0.351 & 0.304 & 0.26 & 0.237 & 0.215\\
  4.50 & 0.331 & 0.283 & 0.238 & 0.214 & 0.192\\
  5.00 & 0.315 & 0.267 & 0.221 & 0.197 & 0.175\\
  6.00 & 0.29 & 0.243 & 0.196 & 0.171 & 0.149\\
  7.00 & 0.271 & 0.225 & 0.179 & 0.154 & 0.132\\
  8.00 & 0.256 & 0.212 & 0.166 & 0.141 & 0.119\\
  9.00 & 0.244 & 0.201 & 0.155 & 0.13 & 0.109\\
 10.00 & 0.234 & 0.192 & 0.147 & 0.122 & 0.101\\
 12.50 & 0.214 & 0.174 & 0.131 & 0.107 & 0.0858\\
 15.00 & 0.199 & 0.161 & 0.12 & 0.0966 & 0.0759\\
 17.50 & 0.187 & 0.151 & 0.112 & 0.0888 & 0.0685\\
 20.00 & 0.178 & 0.143 & 0.106 & 0.0828 & 0.0629\\
 25.00 & 0.163 & 0.131 & 0.096 & 0.0739 & 0.0547\\
 30.00 & 0.152 & 0.121 & 0.0889 & 0.0675 & 0.049\\
 35.00 & 0.143 & 0.114 & 0.0835 & 0.0627 & 0.0447\\
 50.00 & 0.124 & 0.0992 & 0.0724 & 0.0532 & 0.0364\\
 75.00 & 0.105 & 0.0849 & 0.0619 & 0.0445 & 0.0291\\
100.00 & 0.094 & 0.0761 & 0.0555 & 0.0394 & 0.025\\
125.00 & 0.0858 & 0.07 & 0.051 & 0.0359 & 0.0223\\
\hline
\end{tabular}
\caption{Normalized surface temperature for $n=3$\label{tab:SurfTemp}}
\end{table*}

\begin{table*}[H]
\begin{tabular}{llllll}
$(t-t_{\rm peak})/t_0$&$\mathcal{T}/\mathcal{T}_{\rm peak}$~~$(\chi=0.05)$&$\mathcal{T}/\mathcal{T}_{\rm peak}~~(\chi=0.1)$&$\mathcal{T}/\mathcal{T}_{\rm peak}~~(\chi=0.15)$&$\mathcal{T}/\mathcal{T}_{\rm peak}~~(\chi=0.2)$&$\mathcal{T}/\mathcal{T}_{\rm peak}~~(\chi=0.3)$\\
\hline
 -1.75 & 0.111 & 0.134 & 0.116 & 0.0941 & 0.0669\\
 -1.50 & 0.243 & 0.274 & 0.237 & 0.208 & 0.17\\
 -1.25 & 0.455 & 0.457 & 0.429 & 0.405 & 0.369\\
 -1.00 & 0.654 & 0.657 & 0.648 & 0.638 & 0.619\\
 -0.75 & 0.811 & 0.821 & 0.823 & 0.822 & 0.816\\
 -0.50 & 0.919 & 0.927 & 0.93 & 0.931 & 0.93\\
 -0.25 & 0.981 & 0.983 & 0.984 & 0.984 & 0.984\\
 -0.10 & 0.997 & 0.997 & 0.997 & 0.998 & 0.998\\
  0.00 & 1 & 1 & 1 & 1 & 1\\
  0.10 & 0.997 & 0.997 & 0.998 & 0.998 & 0.998\\
  0.25 & 0.981 & 0.984 & 0.985 & 0.985 & 0.985\\
  0.50 & 0.929 & 0.935 & 0.938 & 0.94 & 0.94\\
  0.75 & 0.853 & 0.859 & 0.864 & 0.866 & 0.866\\
  1.00 & 0.768 & 0.768 & 0.77 & 0.772 & 0.771\\
  1.25 & 0.686 & 0.676 & 0.672 & 0.671 & 0.668\\
  1.50 & 0.615 & 0.593 & 0.581 & 0.576 & 0.571\\
  1.75 & 0.557 & 0.525 & 0.505 & 0.497 & 0.488\\
  2.00 & 0.51 & 0.472 & 0.445 & 0.433 & 0.422\\
  2.25 & 0.474 & 0.43 & 0.398 & 0.383 & 0.37\\
  2.50 & 0.444 & 0.397 & 0.361 & 0.344 & 0.33\\
  2.75 & 0.42 & 0.371 & 0.332 & 0.314 & 0.298\\
  3.00 & 0.4 & 0.35 & 0.309 & 0.289 & 0.273\\
  3.50 & 0.37 & 0.318 & 0.273 & 0.252 & 0.235\\
  4.00 & 0.346 & 0.294 & 0.248 & 0.226 & 0.208\\
  4.50 & 0.328 & 0.276 & 0.228 & 0.206 & 0.187\\
  5.00 & 0.313 & 0.262 & 0.213 & 0.19 & 0.171\\
  6.00 & 0.29 & 0.24 & 0.19 & 0.167 & 0.148\\
  7.00 & 0.272 & 0.223 & 0.174 & 0.15 & 0.131\\
  8.00 & 0.258 & 0.211 & 0.162 & 0.138 & 0.119\\
  9.00 & 0.247 & 0.2 & 0.152 & 0.128 & 0.109\\
 10.00 & 0.237 & 0.192 & 0.144 & 0.12 & 0.101\\
 12.50 & 0.217 & 0.175 & 0.129 & 0.105 & 0.0867\\
 15.00 & 0.203 & 0.163 & 0.118 & 0.095 & 0.0767\\
 17.50 & 0.191 & 0.153 & 0.11 & 0.0873 & 0.0694\\
 20.00 & 0.181 & 0.146 & 0.104 & 0.0813 & 0.0637\\
 25.00 & 0.167 & 0.133 & 0.094 & 0.0724 & 0.0554\\
 30.00 & 0.155 & 0.124 & 0.0871 & 0.0661 & 0.0495\\
 35.00 & 0.146 & 0.117 & 0.0818 & 0.0613 & 0.0451\\
 50.00 & 0.127 & 0.102 & 0.071 & 0.0518 & 0.0366\\
 75.00 & 0.109 & 0.0874 & 0.0608 & 0.0431 & 0.0292\\
100.00 & 0.0969 & 0.0785 & 0.0546 & 0.0381 & 0.0249\\
125.00 & 0.0885 & 0.0723 & 0.0503 & 0.0346 & 0.0221\\
\hline
\end{tabular}
\caption{Normalized surface temperature for $n=3/2$ (table headings are the same as in table \ref{tab:SurfTemp}) \label{tab:SurfTemp_n1_5}}
\end{table*}

\section{C. Shock breakout detections by X-ray telescopes}\label{sec:appendix3}
The expected number of detected SN shock breakout events from BSGs compatible with the progenitor of SN1987A, $N_{\rm SNSB}$, is presented for a number of active and non-active X-ray telescopes in table \ref{tab:X-ray telescopes}.
For this analysis, the number of emitted photons for the different progenitors is taken from eq.~\eqref{eq:Ngamma_progenitors} and the number of detections is calculated using eq.~\eqref{eq:Nbreakouts}, assuming $n_{\rm det}=100$ and taking the characteristics of the different telescopes presented in the table.

\begin{table*}[H]
\begin{tabular}{|l|c|c|c|c|c|c|c|}\hline
Instrument & $A_{\rm eff}$\footnote{effective area for 1 keV photons} [$\cm^2$]& $\Omega$ \footnote{Field of view} [${\rm deg}^2$]& $\Phi$\footnote{Point spread function angular size, containing $>50\%$ of the signal of a point source}[${\rm deg}^2$]& $t_{\rm I}$ \footnote{Telescope's lifetime} [yr] & $N_{\rm SNSB,1}^{\rm BSG}$ \footnote{Number of SN1987A-like shock breakout detections in the telescope's lifetime, according to eq.~\eqref{eq:Nbreakouts}, with $E_{\rm in}=0.5\times 10^{51}~\erg$.} & $N_{\rm SNSB,2}^{\rm BSG}$ \footnote{Number of SN1987A-like shock breakout detections in the telescope's lifetime, according to eq.~\eqref{eq:Nbreakouts}, with $E_{\rm in}=10^{51}~\erg$.} & $N_{\rm SNSB,3}^{\rm BSG}$ \footnote{Number of SN1987A-like shock breakout detections in the telescope's lifetime, according to eq.~\eqref{eq:Nbreakouts}, with $E_{\rm in}=2\times 10^{51}~\erg$.} \\ \hline
\textit{Einstein} IPC & 100 & 1.56 & 0.002 & 2.5 & 2.2 & 1.7 & 1.1 \\ \hline
\textit{ROSAT} PSPC & 210 & 3.6 & 0.002 & 9 & 54.5 & 44.1 & 28.8 \\ \hline
\textit{Chandra} ACIS & 615 & 0.07 & 6e-08 & 13 & 7.7 & 6.2 & 4.1 \\ \hline
\textit{Chandra} HRC & 225 & 0.25 & 2e-07 & 13 & 6.1 & 4.9 & 3.2 \\ \hline
\textit{XMM-Newton} EPIC-PN & 1227 & 0.2 & 9e-06 & 13 & 61.8 & 50.0 & 32.6 \\ \hline
\textit{XMM-Newton} EPIC-MOS & 922 & 0.3 & 9e-06 & 13 & 60.4 & 48.8 & 31.9 \\ \hline
\textit{Swift} XRT & 125 & 0.15 & 0.0001 & 8 & 0.9 & 0.7 & 0.5 \\ \hline
\end{tabular}
\caption{SN shock breakout detection with X-ray telescopes \label{tab:X-ray telescopes}}
\end{table*}

\bibliographystyle{apj}

\end{document}